\documentclass[twocolumn,pre,floatfix]{revtex4}

\input psfig.sty
\usepackage{epsfig}
\usepackage{amsmath}
\usepackage{times}

\begin{document}




\title{Integration Schemes for Dissipative Particle Dynamics Simulations: 
       From Softly Interacting Systems Towards Hybrid Models} 

\setcounter{page}{1}

\author{I. Vattulainen}
\address{Laboratory of Physics and Helsinki Institute of Physics, 
Helsinki University of Technology, P.O. Box 1100, FIN--02015 HUT, Finland} 

\author{M. Karttunen} 
\address{Biophysics and Statistical Mechanics Group,
Laboratory of Computational Engineering and Research Center 
for Computational Science and Engineering, 
Helsinki University of Technology, P.O. Box 9400, FIN--02015 HUT, Finland} 

\author{G. Besold} 
\address{Max Planck Institute for Polymer Research, Theory Group, 
P.O. Box 3148, D--55021 Mainz, Germany} 

\author{J. M. Polson} 
\address{Department of Physics, University of Prince Edward Island, 
550 University Avenue, Charlottetown, PEI, Canada C1A 4P3}

\begin{abstract} 

We examine the performance of various commonly used integration 
schemes in dissipative particle dynamics simulations. We consider 
this issue using three different model systems, which characterize 
a variety of different conditions often studied in simulations. 
Specifically we clarify the performance of integration schemes 
in hybrid models, which combine microscopic and meso-scale 
descriptions of different particles using both soft and hard 
interactions. We find that in all three model systems many 
commonly used integrators may give rise to surprisingly pronounced 
artifacts in physical observables such as the radial distribution 
function, the compressibility, and the tracer diffusion coefficient. 
The artifacts are found to be strongest 
in systems, where interparticle interactions are soft and 
predominated by random and dissipative forces, while in systems 
governed by conservative interactions the artifacts are 
weaker. Our results suggest that the quality of any integration 
scheme employed is crucial in all cases where the role of random 
and dissipative forces is important, including hybrid models where 
the solvent is described in terms of soft potentials. Regarding 
the integration schemes, the best overall performance is found 
for integrators in which the velocity dependence of dissipative forces 
is taken into account, and particularly good performance is found 
for an approach in which velocities and dissipative forces are 
determined self-consistently. Remaining temperature deviations from 
the desired limit can be corrected by carrying out the self-consistent 
integration in conjunction with an auxiliary thermostat, in a manner 
that is similar in spirit to the well-known Nos\'e-Hoover thermostat.  
Further, we show that conservative interactions can play 
a significant role in describing the transport properties of simple 
fluids, in contrast to approximations often made in deriving analytical 
theories. In general, our results illustrate the main problems associated 
with simulation methods in which dissipative forces are velocity dependent, 
and point to the need to develop new techniques to resolve these issues.

\end{abstract}

\maketitle


\section{Introduction}

One of the greatest challenges in theoretical physics is to understand 
the basic principles that govern soft condensed matter systems, such 
as polymer solutions and melts, colloidal suspensions, and various 
biological processes. Experimental studies of these complex systems 
are often complemented by numerical simulations of model systems, 
which can provide a great deal of information not easily accessible 
by experiment. In this regard, molecular dynamics \cite{Fre96} (MD) 
is often the method of choice, and indeed it can elucidate
various physical phenomena on a microscopic level. In general, however,
such an atomistic approach is problematic since many intriguing 
processes in soft matter systems are not dictated by microscopic 
details but rather take place at mesoscopic length and time scales 
(roughly 1--1000 nm and 1--1000 ns) which are beyond the practical
limits of MD. In such cases, it is necessary to model soft matter 
systems by viewing them from a larger perspective than from 
a microscopic point of view. In practical terms, this means that 
one has to design ways to simplify the underlying systems as much 
as possible, while still retaining the key properties which are 
expected to govern the processes of interest. Recently, this
approach has attracted wider attention as various
``coarse-grained'' simulation techniques have been developed
\cite{Lad93,Mur98,Mal99,Hoo92,Esp95,Gro97,War98} with the purpose
of studying mesoscopic physical properties of model systems.

Dissipative particle dynamics \cite{Hoo92,Esp95,Gro97,War98} 
(DPD) is particularly well suited for this purpose. DPD is 
characterized by coarse-graining in particle representation, 
which allows studies of systems at mesoscopic length scales
and a simplified description of inter-particle interactions \cite{War98}
allowing for studies at mesoscopic time scales. Since DPD
preserves hydrodynamic modes, one may characterize DPD as 
momentum conserving Brownian dynamics. For these reasons, DPD 
is a very promising method for mesoscopic studies of soft systems 
and recently has attracted considerable interest in studies of polymers,
\cite{War98} microphase separation, \cite{Gro99} and lipid
bilayers, \cite{Ven99} among others.

Despite its advantages, DPD has certain practical problems that 
have to be resolved before extensive use in large-scale simulations. 
Many of them are related to the idea of coarse graining which can 
be done by simplifying molecular representations, and then 
replacing the ``fast'' variables related to the coarse-grained 
degrees of freedom by random noise. The random noise mimics thermal 
fluctuations and hence drives the system. In DPD, this idea is 
implemented by a special ``DPD thermostat'' \cite{Esp95,Gro97,War98} 
in terms of dissipative as well as random pairwise forces such that 
the momentum is conserved. This is a prerequisite for the emergence 
of hydrodynamic flow effects on a macroscopic scale. However, due 
to the DPD thermostat and the resulting stochastic nature of the 
equations of motion, the quest for a suitable integration scheme 
in DPD is a non-trivial task. It has been recently observed 
that various integration schemes commonly used in classical MD 
lead to distinct deviations from the true equilibrium behavior, 
including deviations from the temperature predicted by the 
fluctuation-dissipation theorem, and artificial structures 
observed in the radial distribution function. 
\cite{Gro97,Nov98,Pag98,Gib99,Ott01} These findings demonstrate 
the serious practical problems associated with the use of DPD and 
raise concerns regarding its future application to large-scale 
simulations of soft systems.

A related problem regards hybrid models, where the aim is to 
combine microscopic models of biomolecules with a meso-scale 
modeling of the solvent. \cite{Ahl99,Mal00}  In this promising 
approach, one can examine microscopic properties of complex 
biological molecules in an explicit solvent but with a reduced 
computational cost. While biomolecules are described by hard 
conservative interactions such as Lennard-Jones and Coulombic 
forces, the solvent can be described by DPD as a softly interacting 
fluid. The drawback is that the integration schemes may again 
lead to deviations from the true equilibrium behavior. To our 
knowledge, the role of integration schemes in these cases, where 
both soft and hard interactions are used within a meso-scale DPD 
simulation, has not been studied yet. These examples clearly 
highlight the current need to examine the relative performance of 
different integration schemes in DPD under various conditions, 
and develop new integration techniques where the special features 
of DPD are properly accounted for.

In this work, we study the performance of a number of commonly 
used integration schemes in DPD simulations. They all are based 
on the velocity--Verlet scheme but differ in how the velocity 
dependence of dissipative forces in DPD is taken into account. 
We test the integrators by studying a number of physical observables 
such as the temperature, the compressibility and the tracer diffusion 
coefficient, and evaluate their performance in three different model 
systems. We first examine how the integrators perform in the absence 
of conservative forces. 
This case was partly discussed in our previous work, \cite{Bes00} 
which is here extended by a thorough discussion of the results 
and the details of self-consistent integration schemes suggested 
in Ref.~\onlinecite{Bes00}. 
Then, by increasing the relative importance of conservative 
forces, we eventually obtain a model which is used to assess 
the performance of integration schemes in a hybrid approach.

We find that various commonly used integration schemes in MD and DPD 
indeed lead to pronounced artifacts in actual physical quantities. 
These artifacts are found to be strongest in weakly interacting 
systems, where interactions are soft and dominated by
random and dissipative forces. In the opposite limit, where hard 
conservative interactions govern the system under study, the 
artifacts due to integration schemes are less pronounced. 
We conclude that the quality of an integration 
scheme employed is crucial in all cases where the role of random 
and dissipative forces is important, including hybrid models where 
the solvent is described in terms of soft potentials.

Regarding the integration schemes, best overall performance is 
found for an integration scheme which involves the self-consistent 
determination of particle velocities and dissipative forces. 
For cases where precise temperature control is crucial, we 
further suggest and analyze in detail an additional auxiliary 
thermostat which corrects for the residual temperature deviations.

The outline of the paper is as follows. 
In Sect.~\ref{sec:methods}, we first review the essential
background of DPD and then introduce the three model systems 
studied in this work. The integrators which are tested are described in
Sect.~\ref{sec:integrators}, after which, in Sect.~\ref{Sec:results},
we present and discuss the test results. In Sect.~\ref{sec:diffusion}
we discuss the special case of tracer diffusion behavior of DPD particles 
and compare our findings to previous theoretical descriptions.
Finally, we close this paper with a short
summary and discussion in Sect.~\ref{sec:summary}.

\section{Methods and Models}
\label{sec:methods}

Below we give a short summary of DPD and describe the model 
systems used in this study. For more thorough accounts on DPD, 
see Refs.~\onlinecite{Gro97} and~\onlinecite{War98}.

\subsection{Dissipative Particle Dynamics}

In the present work, we study a simple model fluid system
described by $N$ identical particles each  with mass $m$,
and which have coordinates $\vec{r}_i$, and velocities $\vec{v}_i$.
Interparticle interactions are characterized by the pairwise 
conservative, dissipative, and random forces exerted on 
particle ``$i$'' by particle ``$j$'', respectively, and are 
given by  
\begin{eqnarray}
\label{DPD_forces}
\vec{F}_{ij}^C &=& \alpha\ \omega^{C} (r_{ij})  
   \vec{e}_{ij} , \label{forces:1} \\
\vec{F}_{ij}^D &=& - \gamma\ \omega^{D} (r_{ij}) 
   (\vec{v}_{ij} \cdot  \vec{e}_{ij})\, \vec{e}_{ij} , \label{forces:D} \\
\vec{F}_{ij}^R &=& \sigma \omega^{R} (r_{ij}) \xi_{ij} 
   \vec{e}_{ij} , \label{forces:2}
\end{eqnarray}
where 
$\vec{r}_{ij} \equiv \vec{r}_i - \vec{r}_j$,
$r_{ij} \equiv | \vec{r }_{ij} |$,
$\vec{e}_{ij} \equiv \vec{r}_{ij} / r_{ij}$, and
$\vec{v}_{ij} \equiv \vec{v}_i - \vec{v}_j$.
The $\xi_{ij}$ are symmetric random variables with zero 
mean and unit variance, and are independent for different pairs 
of particles and different times.

The pairwise conservative force of Eq.~(\ref{forces:1}) 
is written in terms of a weight function $\omega^{C} (r_{ij})$,
whose choice is dictated by the system under study. In principle,
$\omega^{C} (r_{ij})$ can include various kinds of forces due to
e.g. electrostatic interactions, as well as descriptions
of detailed intermolecular interactions such as van der Waals 
forces. However, since DPD has been designed to model molecular 
systems on a mesoscopic level, a detailed atomistic description 
of interactions is, in many cases, not necessary. Instead, it is 
often preferable to use soft-repulsive interactions of a relatively 
simple form. This approach is justified 
by observations by Forrest and Suter that coarse graining of 
a molecular representation tends to soften interactions. \cite{For95} 
Recent work by Flekk{\o}y {\it et al.} \cite{Fle99} also supports 
this view. We will return to this issue in Sect.~\ref{Sec:dpdmodel}, 
where the actual form of $\omega^{C} (r_{ij})$ will be discussed 
in more detail.

Unlike the conservative force, the weight functions 
$\omega^{D} (r_{ij})$ and $\omega^{R} (r_{ij})$ of the 
dissipative and random forces cannot be chosen independently.
Physically, $\vec{F}_{ij}^D$ and $\vec{F}_{ij}^R$ 
have to be coupled, since thermal heat generated by the random 
force must be balanced by dissipation. The precise relationship 
between these two forces is determined 
by the fluctuation-dissipative theorem, which sets conditions 
for both the weight functions 
\begin{equation} 
\label{eq:fluc1}
\omega^{D} (r_{ij})  =  [ \omega^{R} (r_{ij}) ]^2 
\end{equation}
and the amplitudes of these forces 
\begin{equation} 
\label{eq:fluc2}
\sigma^2  =  2 \gamma k_B T^{\ast} \,,
\end{equation}
where $T^{\ast}$ is the canonical temperature of the system.

DPD samples phase space according to the canonical ensemble 
with a potential energy determined by the conservative force 
described by Eq.~(\ref{forces:1}). Consequently, static 
properties such as the pair correlation function and the specific 
heat could be calculated equally well using any stochastic 
technique (such as the off-lattice Monte Carlo method). 
On the other hand, as in molecular dynamics, DPD also provides 
a means to calculate dynamical properties of the system. 
Consequently, a method is required to evolve the system in 
time, which in DPD is usually done by integrating Newton's equations
of motion. Unlike the case in standard molecular dynamics, the 
presence of a stochastic contribution to the force in DPD implies
that the equations of motion are now given by the set of stochastic 
differential equations
\begin{eqnarray} 
\label{equs_of_motion}
d \vec{r}_i &=& \vec{v}_i\,dt\,,\label{motion:1}\\
d \vec{v}_i &=& \frac{\displaystyle 1}{\displaystyle m_i}
   \left( \vec{F}_i^{C}\,dt + \vec{F}_i^{D}\,dt + 
   \vec{F}_i^R\,\sqrt{dt}\,\right)\,,\label{motion:2}
\label{equs_of_motion2}
\end{eqnarray}
where \,$\vec{F}_i^{R} = \sum_{j \not= i} \vec{F}_{ij}^{R}$\, 
is the total random force acting on particle ``$i$'', and 
$\vec{F}_i^{C}$ and $\vec{F}_i^{D}$ are defined correspondingly. 
The velocity increment due to the random force in 
Eq.~(\ref{equs_of_motion2}) is written in a form which can be 
given a precise meaning by identifying it as the infinitesimal 
increment of a Wiener process. \cite{Gar83} In practice, 
finite time increments are used in the simulations, and the 
equations of motion 
[Eqs.~(\ref{equs_of_motion}) and (\ref{equs_of_motion2})] have 
to be solved by some integration procedure. As will be seen in 
Sect.~\ref{Sec:results}, this may lead to serious artifacts if 
the special features of DPD are not taken into account.

\subsection{Model systems\label{Sec:dpdmodel}}

In this study, we investigate the performance of various integration 
schemes using three different model systems. They all are based on 
a 3D simple model fluid system with a fixed number of similar 
spherical particles. The differences between the model systems 
arise from interaction effects, which are varied step by 
step from an ideal gas to a more realistic description of 
an interacting fluid system. The models studied here are 
described below.

\subsubsection{Model A} 

We first consider the case characterized by the absence of 
conservative forces ($\alpha = 0$). This choice corresponds 
to an ideal gas (sometimes termed ``ideal DPD fluid'' 
within the framework of DPD), which provides us with some 
exact theoretical results to be compared with those of model 
simulations. Here, the dynamics of the system 
arise only from random noise and the dissipative coupling
between pairs of particles. The random force strength is 
chosen as $\sigma = 3$ in units of $k_B T^{\ast}$, and the 
strength of the dissipative force $\gamma$ is then 
determined by the fluctuation-dissipation relation
in Eq.~(\ref{eq:fluc2}). The weight function 
$\omega^{R} (r_{ij})$ was chosen as in various previous 
works, \cite{Gro97,War98,Gro99,Nov98} 
\begin{equation}
\label{eq:weight}
\omega^{R} (r_{ij}) = \left\{ \begin{array}{ll}
 1 - r_{ij}/r_c & \mbox{\rm , for } r_{ij} \leq r_c \\
 0              & \mbox{\rm , for } r_{ij} > r_c
 \end{array}
 \right .
\end{equation}
where $r_c$ is a cut-off distance. The weight function 
$\omega^{D} (r_{ij})$ is defined via Eq.~(\ref{eq:fluc1}). 
Therefore, the dissipative and random forces are just soft 
pairwise repulsions acting along the line of centers of 
two DPD particles, and $\gamma$ and $\sigma$ are the 
amplitudes which define the maxima of these forces.

In our simulations we use a 3D box of size 
$10 \times 10 \times 10$ with periodic boundary 
conditions, where the length scale is defined by 
$r_c = 1$, and a particle number density $\rho = 4$.

\subsubsection{Model B} 

Model B is a simple interacting DPD fluid. 
Its main difference with respect to model A is the presence of
a conservative force, which we choose to have a strength 
$\alpha = 25$ and a weight function $\omega^{C} (r_{ij})$ of the same 
form as the random force in Eq.~(\ref{eq:weight}). In
all other respects, this model system is identical to model A.

\subsubsection{Model C \label{sec:modelc}}

Model C is a variation of model B. Instead of 
soft potentials, we now use hard conservative interactions. 
The conservative potential between particles ``$i$'' and 
``$j$'' is described by the truncated and shifted 
Lennard-Jones potential 
\begin{equation}
\label{eq:lennardjones}
U_{ij}^C(r_{ij}) = \left\{ 
\begin{array}{ll} 
4\,\epsilon \left[ \left(\frac{\ell}{r_{ij}}\right)^{12} 
                 - \left(\frac{\ell}{r_{ij}}\right)^{6} + \frac{1}{4}
                \right] 
   & ,\, r_{ij} \leq r_c \\
0  & ,\, r_{ij} > r_c 
\end{array}
\right . 
\end{equation}
such that the potential is purely repulsive and decays smoothly
to zero at $r_c$. 
We choose $\ell = 2^{-1/6}$ and $\epsilon = k_B T^{\ast}$, 
and therefore $ r_c = \ell \, 2^{1/6} = 1$. 
For $r_{ij} \geq r_c$, $U_{ij}^C(r_{ij}) = 0$. 
The pairwise conservative force follows directly from 
$\vec{F}_{ij}^C = - \nabla U_{ij}^C$. The dissipative and 
random forces are described by Eq.~(\ref{eq:weight}).

Simulations were carried out in a 3D box of size 
$16 \times 16 \times 16$ with periodic boundary 
conditions, with $\sigma$ ranging from 1 to 200, and 
with particle densities $\rho = 0.1$ and $\rho = 0.7$.

Finally, let us briefly justify the choice of these model systems. 
In model A, the idea is to study integrator-induced 
artifacts in a case, where the role of random noise with respect 
to conservative interactions is as pronounced as possible. Model B 
corresponds to a typical situation where large-scale processes such 
as microphase separation and morphological properties of complex 
systems are studied in terms of coarse-grained particles. 
In that case,  details of molecular representation are no longer 
accounted for, and all interactions are described in terms of 
soft potentials. Finally, model C aims to gauge integrator-induced 
effects in a hybrid approach, where both hard and soft 
interactions are present. One likely scenario of this idea is 
to model solute molecules with realistic atomic force fields, 
while the solvent is described in a coarse grained fashion. 
In the present work we restrict our test simulations to simple 
spherical particles which interact via Lennard-Jones type 
interactions, since that should already allow a reliable 
assessment of integrator-induced artifacts in hybrid models.

\subsection{Practical details} 

One of most important practical aspects within DPD is 
the stochastic nature of the interactions. This is built in to
the random force of Eq.~(\ref{forces:2})
via $\xi_{ij}$, which are independent random
variables with zero mean and unit variance. In the present 
work, we have described them by uniformly distributed random 
numbers $u \in U(0,1)$ such that, for every pair of particles 
at any moment, we generate a different stochastic term 
$\xi = \sqrt{3}(2u - 1)$. This approach is very efficient 
and yields results that are indistinguishable from those 
generated by Gaussian random numbers. \cite{dunweg91a}

In generating the random numbers, we used a pseudorandom 
number generator RAN2, which is based on the 32-bit combination 
generator first proposed by L'Ecuyer \cite{Lec88} and later 
published in Numerical Recipes \cite{Pre92} using shuffling. 
In a recent study, \cite{Vat01}  where several pseudorandom
number generators were tested in DPD model simulations, it was 
demonstrated that RAN2 performs very well in simulations of 
simple fluids.

The length scale in the simulations is defined by 
$r_c = 1$ and the time scale is given in units of 
$r_c \sqrt{ m / k_B T^{\ast}}$. The energy scale is 
defined by setting the desired thermal energy to unity 
via $k_B T^{\ast} = 1$.

The simulations were carried out with particle numbers of 
the order of a few thousand (4000 in models A and B, and 
roughly 2800 in model C for $\rho = 0.7$). The number of 
time steps varied depending on the size of the time 
increment $\Delta t$ such that the total simulation time was 
about 5000 -- 10000 (in units of $r_c \sqrt{ m / k_B T^{\ast}}$).

\section{Integrators}
\label{sec:integrators}

One of the central issues in molecular dynamics calculations 
is the integration of the equations of motion. In the  context 
of MD, the present understanding of this issue is rather clear
and comprehensive. \cite{Tuc00}  However, in the case of
DPD simulations the situation is more problematic. To 
clarify this, let us consider the equations of motion in 
detail. Using Eq.~(\ref{motion:2}) for the velocity term we obtain
\begin{eqnarray*} 
d \vec{v}_i &=& \frac{\displaystyle 1}{\displaystyle m_i}
   \left[ 
   \alpha \, dt \, \sum_{j \not= i} \omega^{C} (r_{ij}) \, \vec{e}_{ij} 
   \, + \right. \\
   & & - \, \gamma \, dt \, \sum_{j \not= i}  
   \omega^{D} (r_{ij}) (\vec{v}_{ij} \cdot 
   \vec{e}_{ij})\, \vec{e}_{ij} \, + \\
   & & \left. 
   + \, \sigma \, \sqrt{dt}\, 
   \sum_{j \not= i} \omega^{R} (r_{ij}) \, \xi_{ij} \, \vec{e}_{ij} 
   \right] \,, 
\end{eqnarray*}
which immediately reveals two potential problems.
First, due to the stochastic nature of interactions,
the time reversibility is no longer guaranteed.
Another serious problem arises from the dissipative 
forces, which depend on the pairwise velocities of all pairs 
of particles. This seemingly minor detail is absent from classical 
MD simulations but leads to significant problems in DPD simulations, 
including artifacts in various physical quantities measured from 
simulation studies. \cite{Gro97,Nov98,Pag98,Gib99}  In principle, 
this problem could be solved by finding a self-consistent solution 
for both dissipative forces and particle velocities by inverting an 
appropriate interaction matrix of size $N \times N$ at every time 
step. However, it is obvious that this approach is generally not feasible,
and thus one must search for more practical solutions.

\subsection{Simple velocity-Verlet based integration schemes}

We use the velocity-Verlet scheme \cite{Ver67} 
as a starting point and consider various previously used 
integrators based on this approach. These are summarized in 
Table~\ref{table1}, where the acronym ``MD--VV'' corresponds 
to the standard velocity-Verlet algorithm used in classical 
MD simulations. The MD--VV scheme is (in the case of solely 
conservative forces) a time-reversible and symplectic second-order 
integration scheme, which has been shown to be relatively 
accurate in typical MD simulations especially at large time 
steps. \cite{All93}  Although some higher-order algorithms 
are more accurate than the MD--VV, \cite{Mar95} 
the simplicity of MD--VV makes it a good starting point 
for further development.

Unlike in molecular dynamics, the forces in DPD depend on 
the velocities. This fact is not accounted for within 
the MD--VV scheme. In an attempt to deal with this complication, 
Groot and Warren subsequently proposed
\cite{Gro97} a modified velocity-Verlet integrator
[``GW($\lambda$)'' in Table~\ref{table1}]. In this approach, 
the forces are still updated only once per integration step, 
but the dissipative forces are evaluated based on intermediate 
``predicted'' velocities $\vec{v}_{i}^{\,\circ}$. The underlying 
idea of $\vec{v}_i^{\,\circ}$ is to use a phenomenological 
tuning parameter $\lambda$, which mimics higher-order corrections 
in the integration procedure. The case $\lambda = 1/2$ 
corresponds to the usual MD--VV, while other 
suggested choices range from zero to one. The problem is that 
the relative merits of different numerical values for $\lambda$ 
are poorly understood. 
Groot and Warren studied the temperature control in an interacting 
fluid and found that $\lambda = 0.65$ works better than 
$\lambda = 1/2$. \cite{Gro97} In a different study, Novik and 
Coveney concluded that $\lambda = 1/2$ gives a more accurate 
temperature than $\lambda = 1$. \cite{Nov98}  Thus, it is
evident that the optimal value of $\lambda$, which minimizes
the temperature shift and other artifacts, depends on model
parameters and has to be determined empirically.

Recently, Gibson {\it et al.} proposed \cite{Gib99} 
a slightly modified version of the GW integrator. This 
``GCC($\lambda$)'' integrator updates the dissipative 
forces [step (5) in Table~\ref{table1}] for a second time 
at the end of each integration step. This approach 
suffers from the same problem as the GW integration scheme, 
i.e., it uses a phenomenological parameter whose 
optimization depends on the system and the conditions 
that are being modeled. Based on a few model studies by 
Gibson {\it et al.}, values of $\lambda$ between $1/2$ 
and 1 may be preferable to smaller values. \cite{Gib99}

Despite the use of a phenomenological parameter, the GCC scheme 
is a promising approach for DPD simulations. A rational starting 
point is to fix $\lambda$ to a value of $1/2$, which leads to 
an integrator equivalent to the MD--VV scheme supplemented by 
the second update of the dissipative forces. This Verlet-type 
integrator, here termed ``DPD--VV'', is particularly appealing 
because it does {\it not} involve any tuning parameters, yet it 
takes the velocity-dependence of the dissipative forces at least 
approximately into account. In addition, it is computationally 
very efficient since the additional update of dissipative forces 
is an easy task compared to the time-consuming part of updating 
neighbor tables.

In this work, we consider besides the schemes 
 GW$(\lambda = 1/2) = $ MD--VV and 
GCC$(\lambda = 1/2) = $ DPD--VV also  
 GW$(\lambda = 0.65)$ and GCC($\lambda = 0.65$).

\subsection{Self-consistent velocity-Verlet integrators}

Unfortunately, as will be shown in Sect.~\ref{Sec:results}, 
all of the above integrators display pronounced unphysical 
artifacts in the radial distribution function $g(r)$, and thus 
do not produce the correct equilibrium properties (see results 
and discussion below). This highlights the need for an 
approach in which the velocity-dependence of dissipative 
forces is fully taken into account. In principle this 
problem can be easily addressed by solving the velocities 
and dissipative forces in a self-consistent fashion. In 
practice, however, there is no unique way to do this. In 
this work, we present in Table~\ref{table2} the update schemes 
for two self-consistent schemes which are variants of DPD--VV. 
The basic variant SC--VV, which is similar in spirit to the 
self-consistent leap-frog scheme introduced by Pagonabarraga 
{\it et al.}, \cite{Pag98} determines the velocities and 
dissipative forces self-consistently through functional 
iteration, and the convergence of the iteration process is 
monitored by the instantaneous temperature $k_B T$.

In the second approach, which we call SC--Th, we couple the 
system to an auxiliary thermostat and obtain an 
``extended-system'' method in the spirit of Nos{\'e}-Hoover. 
\cite{Thi99} The idea behind this approach is that whenever 
$\langle k_B T \rangle$ deviates from $k_B T^{\ast}$, the 
dissipation rate is on average not balanced by the excitation 
rate (due to the stochastic forces) in the system. Here we 
attempt to correct this imbalance by ``fine-tuning'' the 
dissipation rate by an auxiliary thermostat. In order to 
preserve the pairwise conservation of momentum in DPD, 
this auxiliary thermostat is implemented by employing 
a {\it fluctuating} dissipation strength, defined by 
\begin{equation} 
\label{thermostat}
\gamma(t) = \frac{\sigma^2}{2 k_B T^{\ast}} ( 1 + \eta(t)\,\Delta t) ,
\end{equation} 
where $\eta$ is the thermostat variable. The rate of change of 
$\eta$ is proportional to the instantaneous temperature deviation 
$\dot{\eta} = C (k_B T - k_B T^{\ast})$ where $C$ is a coupling 
constant, step (i) in Table~\ref{table2}. This first-order 
differential equation must be integrated [step (ii)] 
simultaneously with the equations of motion. In this respect 
our thermostat resembles the Nos\'e{}-Hoover thermostat 
familiar from MD simulations. \cite{Thi99} Equation~(\ref{thermostat}) 
can be interpreted as an expansion of the optimal $\gamma$ in 
terms of $\Delta t$ up to the linear order. This ansatz ensures 
that the correct continuum version of DPD is regained for 
$\Delta t \rightarrow 0$. Also note that the coupling constant 
$C$ has to be chosen with care. Very small values of $C$ require 
considerably longer simulation times, while too high values may 
bias the temperature distribution as well as the transport
coefficients. For the simulations reported here, we optimized
$C$ by studying the characteristic decay time of 
$\langle \gamma(t) \gamma(0) \rangle$. In this manner, we ensured
that the chosen time scale of the dissipation strength
fluctuations did not interfere with the underlying dynamics 
of the system (in the absence of auxiliary thermostat).

\section{Performance of integrators}
\label{Sec:results}

\subsection{Physical quantities studied}

We characterize the integrators by studying a number of 
physical observables. After equilibrating the system, we first 
calculate the average kinetic temperature $\langle k_B T \rangle$, 
whose conservation is one of the main conditions for reliable 
simulations in the canonical ensemble. Next, we consider the 
radial distribution function $g(r)$, \cite{Boo80}  which is 
one of the most central observables in studies of liquids and 
solid systems. For the ideal gas (model A), the radial 
distribution function provides an excellent test for the 
integrators, since then $g(r) \equiv 1$ in the continuum 
limit. Therefore, any deviation from unity has to be interpreted 
as an artifact due to the integration scheme employed. For the
other models there are no such straightforward theoretical
predictions. Consequently, we test each model by comparing
the results of different integrators to one another.

Artifacts in $g(r)$ are also reflected in the relative 
isothermal compressibility 
\begin{equation}
{\widetilde\kappa}_T\equiv\kappa_T / \kappa_T^{\mathrm{\it ideal}} , 
\end{equation}
where $\kappa_T^{\mathrm {\it ideal}} = (\rho\,k_B T^{\ast})^{-1}$ 
denotes the compressibility of the ideal gas in the continuum limit. 
For an arbitrary fluid, ${\widetilde \kappa}_T$ is related to $g(r)$ by 
\begin{equation}
{\widetilde \kappa}_T = 1 + 4\pi\rho \int_0^{\infty} dr \, r^2 \, [g(r) - 1], 
\end{equation}
and thus any deviation from ${\widetilde \kappa}_T = 1$ for the 
ideal gas (model A) indicates an integrator-induced artifact. 
For models B and C, ${\widetilde \kappa}_T $ serves as a measure 
of integrator-induced artifacts after a thorough comparison of 
results of different integrators relative to each other.

To gauge the underlying problems in the {\it dynamics} 
of the system, we consider the tracer diffusion coefficient 
\begin{equation} 
D_T = \lim_{t \to \infty} 
\label{Eq:msd}
\frac{1}{6Nt} \sum_{i=1}^{N} 
\langle [ \vec{r}_i(t) - \vec{r}_i(0) ]^2 \rangle , 
\end{equation}
in which the mean-square displacement 
$\langle [ \vec{r}_i(t) - \vec{r}_i(0) ]^2 \rangle $ 
is the average squared distance that the tagged particle 
travels during a time interval $t$. In the long-time limit 
one obtains the tracer diffusion coefficient $D_T$, which 
characterizes the distance ${\ell_D} \sim \sqrt{D_T \, \delta t}$ 
travelled by a particle during a long time period $\delta t$.

Another way to gauge the effects of the numerical integration methods
on dynamical quantities is to monitor the velocity-correlation function
\begin{equation} 
\label{Eq:phit}
\phi(t) = \frac{1}{N} \sum_{i=1}^{N} 
\langle \vec{v}_i(t+t') \cdot \vec{v}_i(t') \rangle , 
\end{equation}
which defines the tracer diffusion coefficient 
through the Green-Kubo formula \cite{Boo80}
\begin{equation} 
D_T = \frac{1}{3} \int_{0}^{\infty} dt\, \phi(t) . 
\end{equation}

We note that Eqs.~(\ref{Eq:msd}) and (\ref{Eq:phit}) are
complementary approaches for testing the integrators. First, the 
tracer diffusion coefficient can easily be measured from simulations 
via Eq.~(\ref{Eq:msd}), and it provides a way to characterize 
how possible deviations from the true dynamical behavior 
accumulate together. On the other hand, the velocity 
correlation function $\phi(t)$ provides relevant 
information of the short-time dynamics of the tagged particle, 
prior to the region where Eq.~(\ref{Eq:msd}) becomes well defined. 
As an example, the leading term $\phi(0)$ provides information about
temperature conservation, since for fluid systems 
$\phi(0) = \langle k_B T \rangle / 3m$. In addition, since the 
definition of $D_T$ requires $\phi(t)$ to decay to zero
at long times, the decay of $\phi(t)$ can be used to characterize 
possible shortcomings in the dynamics of the system.

\subsection{Results for model A} 
\label{sec:res_modela}

\begin{figure}[htb]
\mbox{\hspace{1.4cm}}
\psfig{file=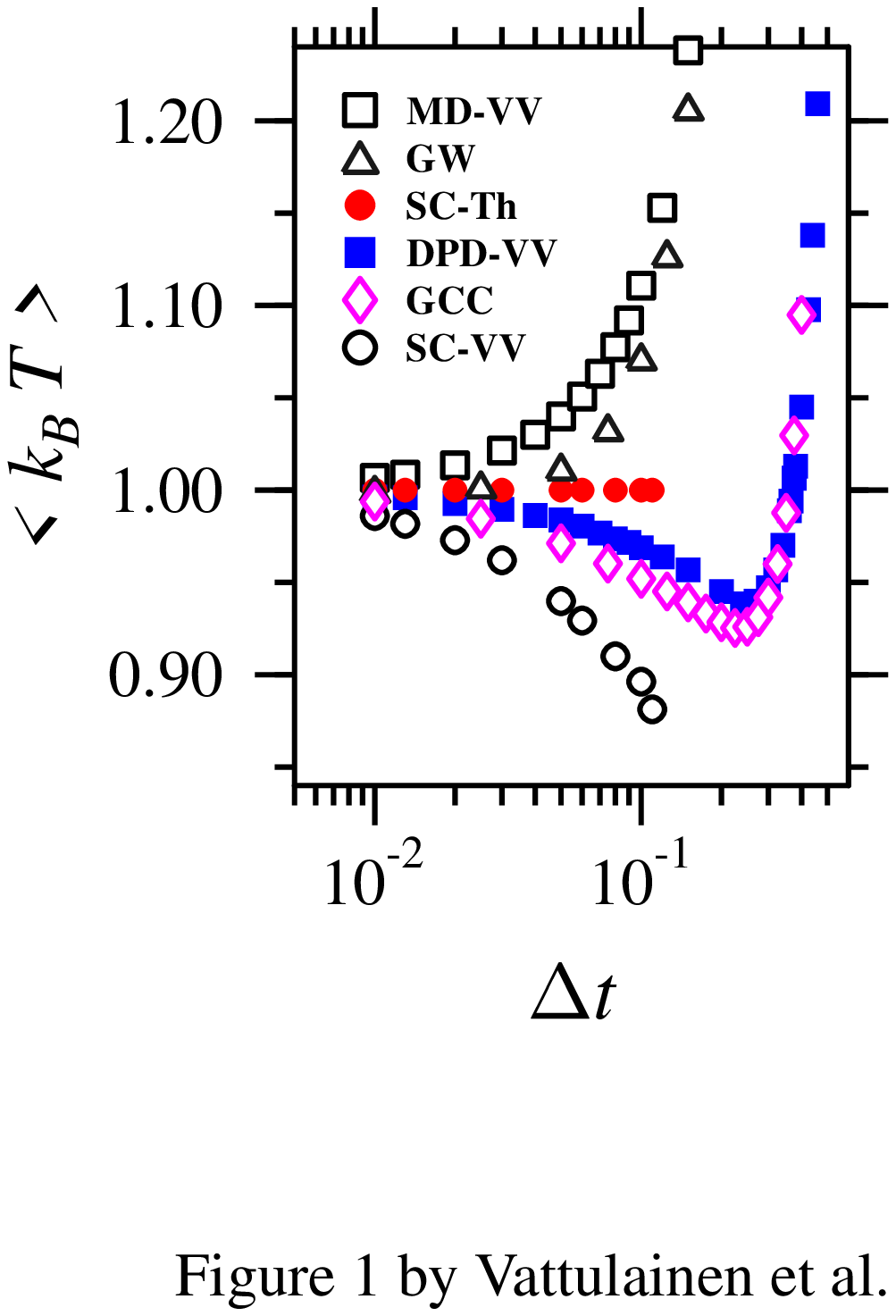,width=5.0cm,bbllx=70pt,bblly=120pt,bburx=400pt,bbury=500pt,clip=}
\caption{\label{fig1}
Results for the deviations of the observed temperature 
$\langle k_B T \rangle$ from the desired temperature 
$k_B T^{\ast} \equiv 1$ vs. the size of the time step $\Delta t$ 
in model A. Results of GW and GCC are for $\lambda = 0.65$. 
}
\end{figure}

First, we discuss the deviations of the observed kinetic temperature
$\langle k_B T \rangle$ from the DPD-thermostat temperature $k_B T^{\ast}$.
For MD--VV this temperature shift, shown in Fig.~\ref{fig1}, is
always positive and increases monotonically with $\Delta t$. For 
DPD--VV, $\langle k_B T \rangle$ first decreases with increasing 
$\Delta t$, then exhibits a minimum at $\Delta t \approx 0.25$, 
and eventually 
becomes larger than $k_B T^{\ast}$. The self-consistent approach 
SC--VV exhibits a negative, monotonically increasing temperature 
shift up to $\Delta t \approx 0.13$, where this scheme becomes
unstable at the employed particle density.
Most importantly, and perhaps most surprisingly, we find
that the modulus of the temperature 
deviation is even larger than 
for DPD--VV. This finding 
contrasts with the findings of a recent study by Pagonabarraga {\it et al.},
\cite{Pag98}  who studied the 2D ideal gas using a self-consistent 
version of the leap-frog algorithm, and found good temperature 
control for $\Delta t = 0.06$ at $\rho = 0.5$. This discrepancy 
can be explained by our observation for the 3D ideal gas that 
the temperature shift is in general more pronounced at higher
densities. A similar effect is found, if the strength of the
interactions is increased. This suggests that temperature
deviations and other related problems due to large time 
steps become more pronounced when the role of interparticle 
interactions is enhanced.

In cases where temperature preservation is crucial in calculating
equilibrium quantities, the self-consistent scheme with an auxiliary  
thermostat, SC-Th, is clearly the method of choice, as is evident from
the results shown in Fig.~\ref{fig1}. For this extended-system method, 
we find that the temperature deviations diminish by over two orders of 
magnitude, with a modulus typically of the order of $10^{-5}$ to $10^{-4}$.
The auxiliary thermostat thus performs very well as long as the 
iteration procedure within the self-consistent scheme remains stable.

\begin{figure}[htb]
\mbox{\hspace{0.0cm}}
\psfig{file=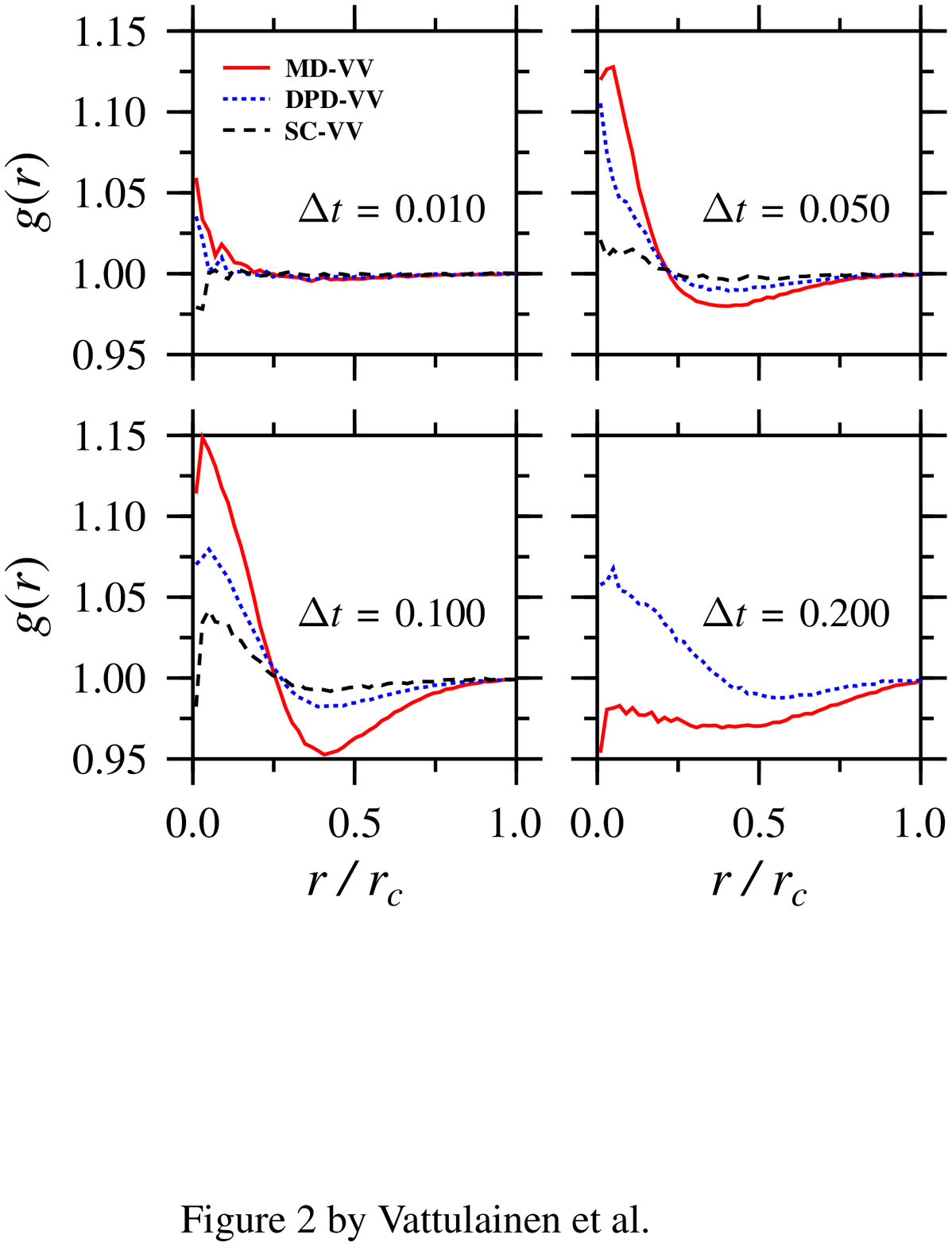,width=9.0cm,bbllx=20pt,bblly=200pt,bburx=600pt,bbury=660pt,clip=}
\caption{\label{fig2}
Radial distribution functions $g(r)$ vs. $\Delta t$ in model A 
for the integration schemes MD--VV, DPD--VV, and SC--VV. Results 
of GW and GCC are almost similar to MD--VV and DPD--VV, 
respectively, and are therefore omitted here. 
}
\end{figure}

Results for $g(r)$ are shown in Fig.~\ref{fig2}. We find 
that the deviations from the ideal gas limit $g(r) = 1$ are very 
pronounced for MD--VV, indicating that even for small time 
steps this integration scheme gives 
rise to unphysical correlations. The performance of DPD--VV
is clearly better, while the self-consistent scheme SC--VV 
leads to even smaller deviations. For the self-consistent
scheme with an auxiliary thermostat SC--Th, we found virtually 
the same results for $g(r)$ as for the self-consistent scheme 
without the thermostat. The results for GW and GCC 
(for a few values of $\lambda$) integrators  
were approximately the same as those of MD--VV
and DPD--VV, respectively. 
For all integrators, the artificial structure in $g(r)$ 
typically becomes more pronounced with increasing time increment 
$\Delta t$. It is noteworthy that the bias introduced 
by the self-consistent integrators for $\Delta t = 0.10$
is comparable to that introduced by MD--VV already for
$\Delta t = 0.01$.

\begin{figure}[htb]
\mbox{\hspace{0.0cm}}
\psfig{file=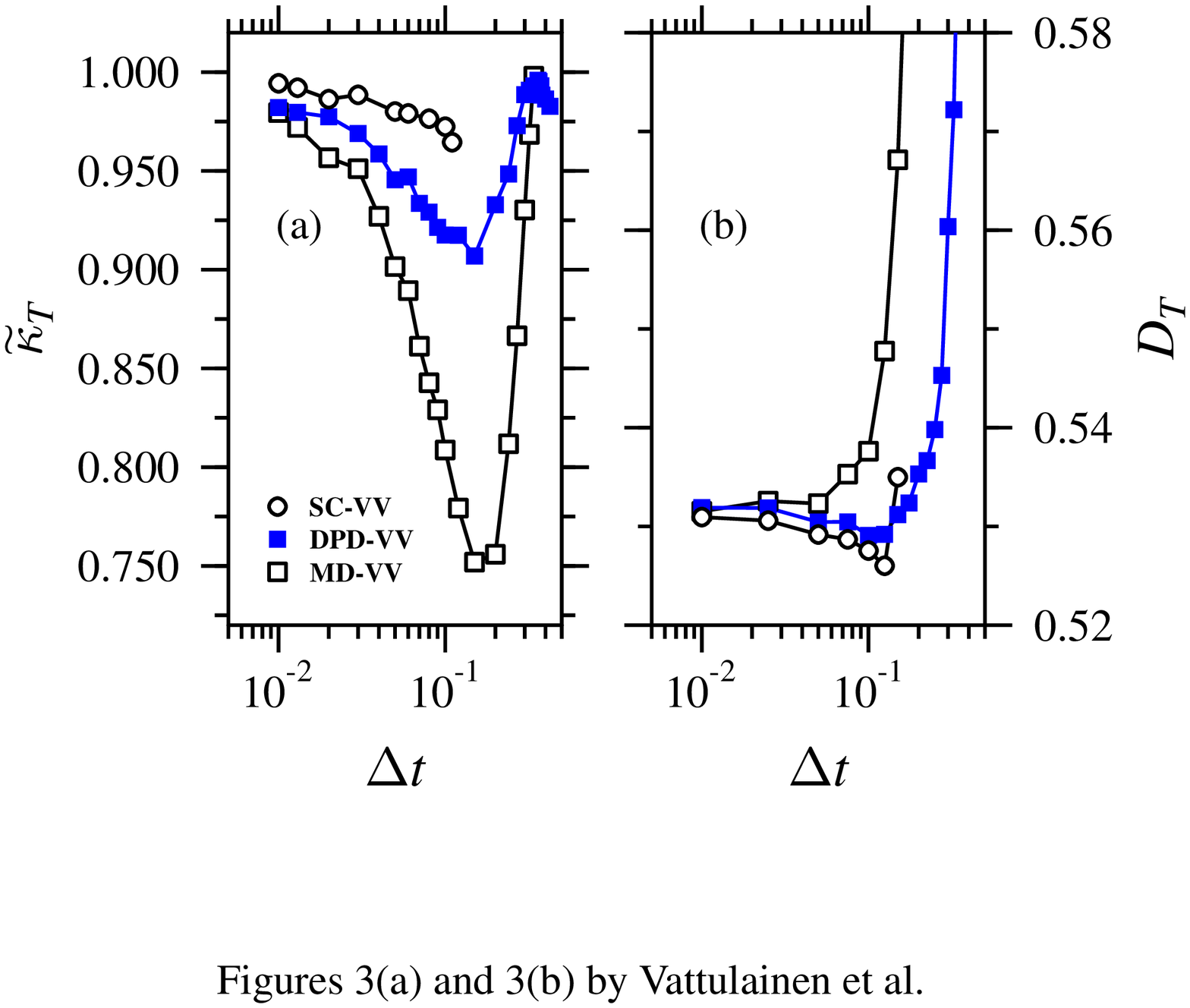,width=9.0cm,bbllx=20pt,bblly=120pt,bburx=600pt,bbury=550pt,clip=}
\caption{\label{fig3}
(a) The relative isothermal compressibilities $\widetilde{\kappa}_T$ 
vs. $\Delta t$ evaluated from $g(r)$ in model A. Ideally one would 
obtain $\widetilde{\kappa}_T = 1 $, and the deviations from this limit 
reflect artifacts due to the integration procedure. 
(b) Results for the tracer diffusion coefficient $D_T$ vs. the 
time step $\Delta t$ in model A. 
}
\end{figure}

The relative isothermal compressibilities $\widetilde{\kappa}_T$ 
evaluated from $g(r)$ are shown in Fig.~\ref{fig3}(a). 
The best performance is found for the self-consistent integrators 
SC--VV and SC--Th, whose behavior is essentially similar, and 
for the DPD--VV whose results are almost equally good. 
In general, the qualitative behavior of $\widetilde{\kappa}_T$ 
reflects our findings for $g(r)$. \cite{compr} The magnitude 
of deviations from $\widetilde{\kappa}_T = 1 $ is astounding, 
however, and raises serious concern for studies of response 
functions such as the compressibility for interacting fluids 
close to phase boundaries. Similarly, the results for tracer 
diffusion in Fig.~\ref{fig3}(b) indicate that DPD--VV and the 
self-consistent approach SC--VV work well up to reasonably 
large time steps, while the other integrators were found to 
perform less well. Further studies regarding the decay of the 
velocity correlation function $\phi(t)$ gave similar conclusions, 
although the size of the artifacts in tracer diffusion are best 
demonstrated by $D_T$. Nevertheless, the decay of velocity 
correlations in tracer diffusion is sensitive to the choice 
of the integrator.

We now discuss some more general aspects concerning the performance of the
self-consistent integrator SC--Th. Based on our results
for $\langle k_B T \rangle$, $g(r)$, and $\widetilde{\kappa}_T$, 
the auxiliary thermostat performs very well. This provides 
clearcut evidence that the SC--Th scheme is useful for studies 
of equilibrium quantities such as the specific heat, which are 
determined by the conservative forces and which are not 
influenced by the details of the dynamics.
However, it  is less clear whether the SC--Th scheme is 
useful for studies of dynamical quantities. To illustrate 
this point, let us consider the motion of a Brownian particle 
as an example. It is characterized by the Langevin equation 
\begin{equation} 
M \frac{d \vec{v}(t)}{dt} = - M \eta \, \vec{v}(t) + \vec{F}(t)  , 
\end{equation}
where $M$ is the mass of the Brownian particle and $\eta$ 
is the friction coefficient which reflects dissipative forces. 
The remaining random term $\vec{F}(t)$ is the driving force due 
to collisions with the solvent particles, whose mass is negligible 
compared to $M$. In this case, one finds that \cite{Han86} 
\begin{equation} 
\label{eq:dt_brownian_eta} 
D_T =  \frac{k_B T}{M\eta} \sim \frac{1}{\eta} , 
\end{equation}
which serves to demonstrate that the tracer diffusion of 
a Brownian particle is clearly affected by the dissipation 
rate. Although the motion of DPD particles is not equivalent 
to Brownian motion, the two cases are related. This
example highlights how any change in the dissipation 
properties may affect diffusion behavior. This problem  
could arise within the SC--Th scheme, since there the strength 
of dissipation is not fixed but it fluctuates in time. Clearly, 
the significance of this issue has to be examined in detail.

\begin{figure}[htb]
\mbox{\hspace{0.0cm}}
\psfig{file=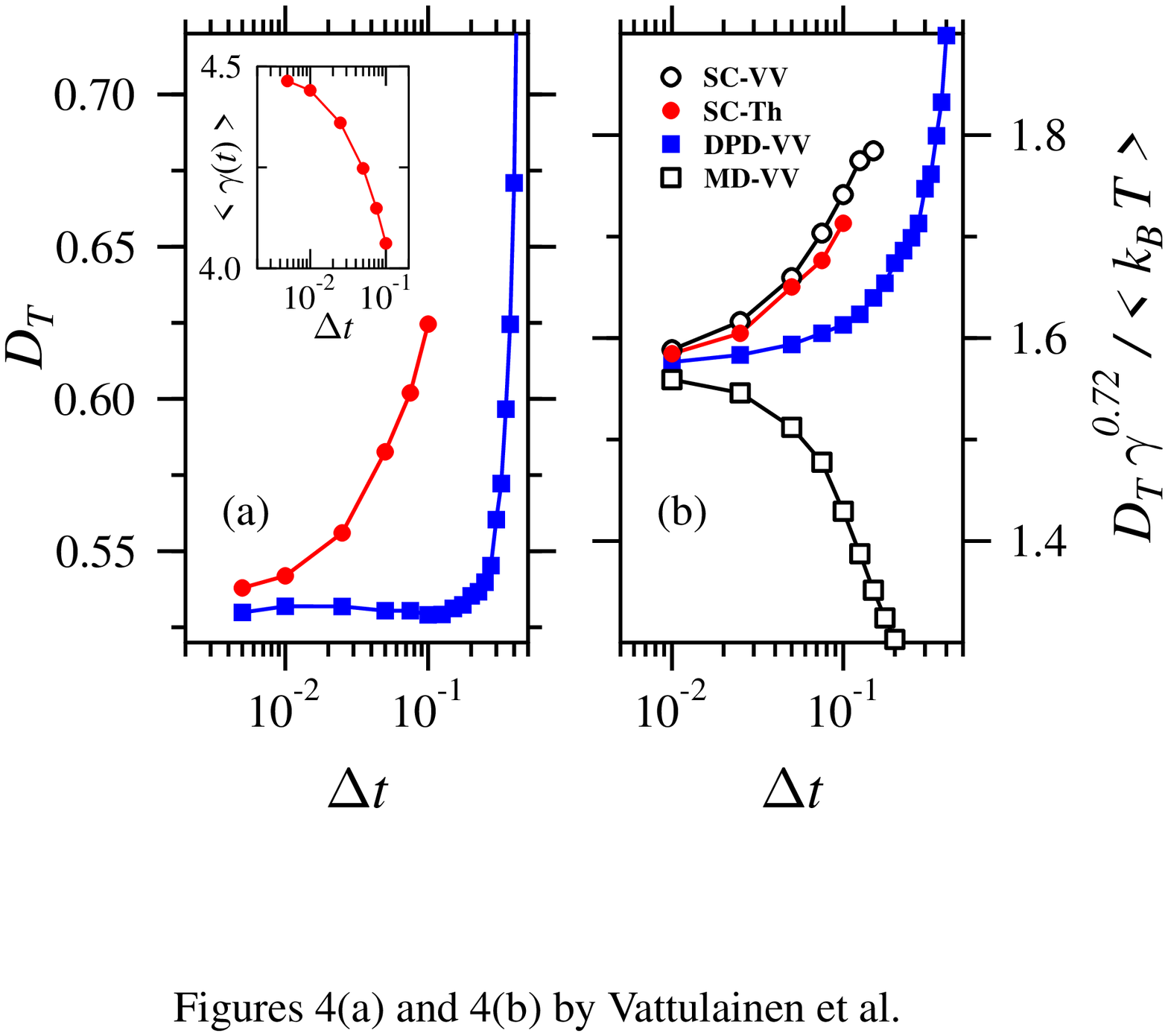,width=9.0cm,bbllx=20pt,bblly=120pt,bburx=600pt,bbury=550pt,clip=}
\caption{\label{fig4}
(a) Results for the tracer diffusion coefficient $D_T$ vs.  
    $\Delta t$ in model A for the SC--Th integrator, in 
    which $\gamma$ is not fixed but fluctuates in time. 
    Results of DPD--VV are also given for the purpose of 
    comparison. The inset illustrates the dependence of 
    $\langle \gamma(t) \rangle$ on $\Delta t$ 
    for the SC--Th integration scheme. 
(b) The scaled tracer diffusion coefficient 
    $D_T \gamma^x / \langle k_B T \rangle$ with $x = 0.72$. 
}
\end{figure}

In Fig.~\ref{fig4}(a) we show the tracer diffusion coefficient 
for the SC--Th integrator versus the size of the time step 
$\Delta t$. In the inset is shown the average strength of 
the dissipative force $\langle \gamma(t) \rangle$ as a function 
of $\Delta t$. The results reveal that $D_T$ converges to the 
correct limit at small $\Delta t$, which is expected since 
$\langle \gamma(t) \rangle \rightarrow \sigma^2 / 2 k_B T^{\ast}$
as  $\Delta t \rightarrow  0 $. 
For larger time steps, $D_T$ clearly deviates from the correct 
behavior. This is due to temperature deviations 
($\langle k_B T \rangle < k_B T^{\ast}$) 
within the {\it original}\, SC--VV scheme (without an auxiliary 
thermostat). These deviations are corrected by the auxiliary 
thermostat by decreasing the average dissipation rate, which 
in turn increases the diffusion rate.

From the discussion above, it is clear
that the dissipation rate within SC--Th depends
on $\Delta t$. Consequently, the transport properties may not be
properly described if temperature deviations due to the 
self-consistent iteration procedure are too large. 
In the present model, this implies that a direct comparison of 
diffusion properties between SC--Th and other integrators is not 
meaningful. For the purpose of completeness, however, let us 
compare their properties in a slightly modified fashion. Instead 
of comparing the diffusion coefficients themselves, we compare their 
scaled counterparts $D_T \, \gamma^{x} / \langle k_B T \rangle$. 
This idea is based on an ansatz that tracer diffusion within 
DPD can be written as 
\begin{equation} 
\frac{D_T}{\langle k_B T \rangle} \sim \left( \frac{1}{\gamma} \right)^x . 
\end{equation} 
In Brownian motion, with $\gamma$ substituted for $\eta$ 
in Eq.~(\ref{eq:dt_brownian_eta}), we find the exponent $x = 1$. 
For DPD, the situation is different and one finds the behavior 
of $x$ to be more complex (see Sect.~\ref{sec:diffusion} and 
Fig.~\ref{fig9} for further discussion). 
Detailed studies under present conditions with 
the DPD--VV scheme (with small $\Delta t$) gave 
$D_T / \langle k_B T \rangle \sim 1 / \gamma^{0.72}$. 
When this dependence on the 
dissipation strength is taken into account, we obtain the results
shown in Fig.~\ref{fig4}(b). Obviously the SC--Th scheme now works
better, but is nevertheless still less accurate than
DPD--VV, for example. This finding simply demonstrates that 
any change in dissipation may lead to further changes in the 
dynamic behavior and should be taken into account in the use of 
auxiliary thermostats. For this reason, we feel that the SC--Th 
scheme is not an ideal approach for studies of dynamical 
quantities by DPD.

Problems of a similar nature are faced in MD studies with the 
Nos\'e-Hoover thermostat, in which the temperature of the system 
is controlled by a ``thermodynamic friction coefficient'' which 
is allowed to evolve in time. \cite{Fre96}  Thus the
present problem with SC--Th is not specific to DPD simulations. 
Furthermore, as will be seen in Sect.~\ref{sec:resultsmodelb}, 
the SC--Th scheme works quite well even for dynamical quantities,
when conservative interparticle interactions are included in 
the model.

\subsection{Results for model B} 

\label{sec:resultsmodelb}

We next consider model B, which describes a fluid with
relatively strong but soft conservative interactions. This situation 
is often met in DPD simulations of polymer dynamics and phase 
separation, among others. Clearly, it is important to understand the
effects of the integrators on the results in these cases.

\begin{figure}[htb] 
\mbox{\hspace{1.4cm}}
\psfig{file=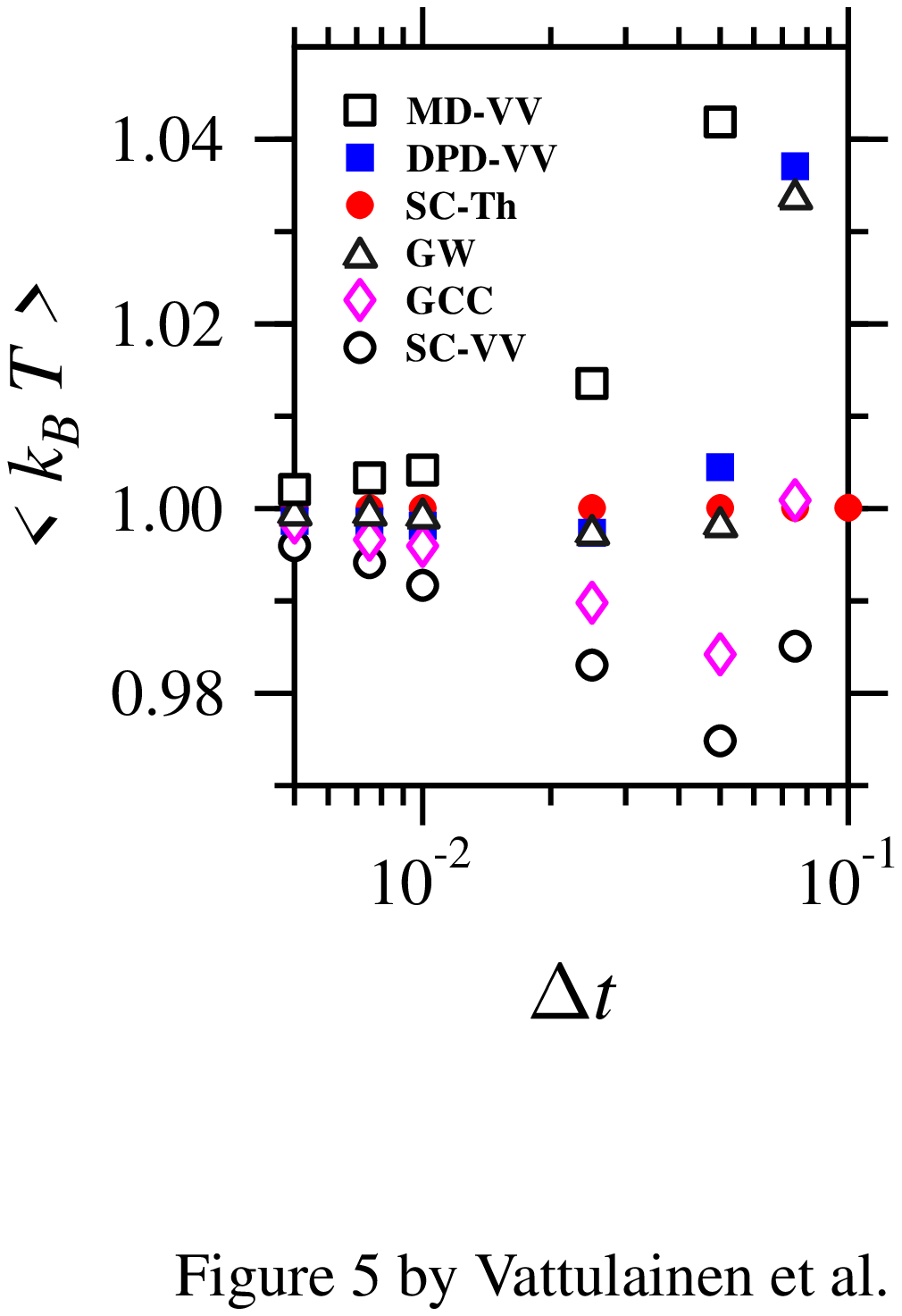,width=5.0cm,bbllx=70pt,bblly=120pt,bburx=400pt,bbury=500pt,clip=}
\caption{\label{fig5}
Results for the deviations of the observed temperature 
$\langle k_B T \rangle$ from the desired temperature 
$ k_B T^{\ast} = 1 $ vs. the size of the time step $\Delta t$ 
in model B. Results of GW and GCC are for $\lambda = 0.65$. 
}
\end{figure}

As a first and demonstrative topic, we again start by 
considering the deviations of the observed actual temperature 
$\langle k_B T\rangle$ from the desired temperature 
$k_B T^{\ast}$. Results shown in Fig.~\ref{fig5} for model B 
reveal that the behavior of the integrators is very similar 
to that found for model A in Sect.~\ref{sec:res_modela}. 
The largest temperature deviations are found for MD--VV and SC--VV, 
and the artifacts due to GCC are almost equally pronounced. 
The performances of DPD--VV and GW are better, while
the SC--Th scheme is found to be superior to all of them.

\begin{figure}[htb] 
\mbox{\hspace{0.0cm}}
\psfig{file=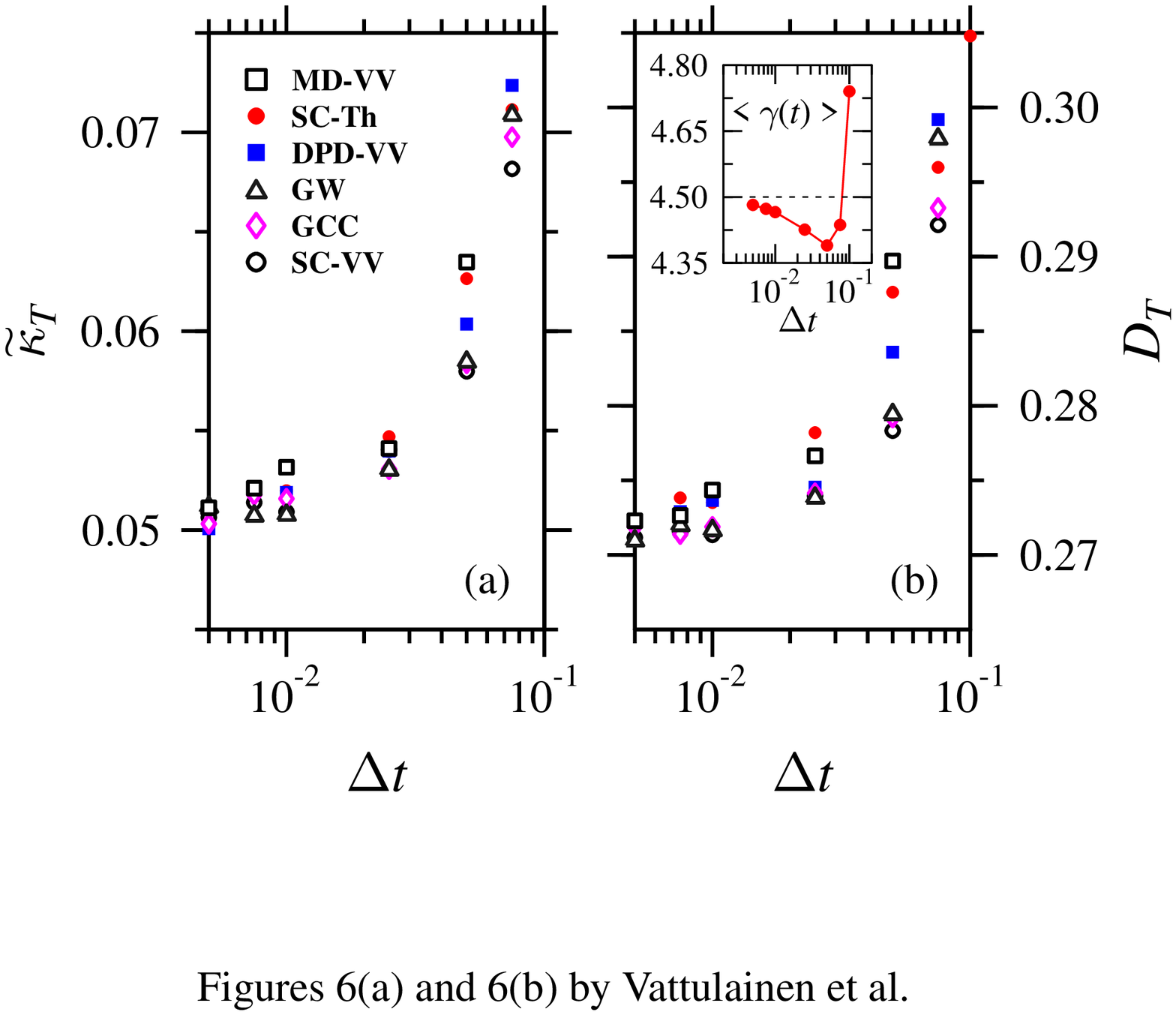,width=9.0cm,bbllx=20pt,bblly=120pt,bburx=600pt,bbury=550pt,clip=}
\caption{\label{fig6}
(a) The relative isothermal compressibilities 
    $\widetilde{\kappa}_T$ evaluated from $g(r)$ in model B. 
(b) Results for the tracer diffusion coefficient $D_T$ 
    vs. $\Delta t$ in model B. Results of GW and GCC are 
    for $\lambda = 0.65$. The inset illustrates the 
    dependence of $\langle \gamma(t) \rangle$ on $\Delta t$ 
    for the SC--Th integration scheme as compared to 
    $\gamma = 4.5$ determined by the fluctuation-dissipation 
    theorem.
}
\end{figure}

Results for the radial distribution function $g(r)$ resemble those 
for any simple interacting fluid, in this case with a minor peak 
at $r \approx 0.86 \, r_c$, and another smaller one around 
$r \approx 1.55 \, r_c$. The radial distribution functions
of different integrators were essentially similar (data not
shown). Thus, it is not too surprising that the compressibility 
data shown in Fig.~\ref{fig6}(a) do not reveal major differences 
between different integrators. The results of all integrators 
are the same to within $\pm 1$\% for $\Delta t \leq 0.01$.
The differences between different integrators are very clear 
only at relatively large time 
steps, where the MD--VV is found to be the poorest and the SC--VV 
the best integration scheme of the ones considered here. 
The results for tracer diffusion in Fig.~\ref{fig6}(b) 
support these conclusions.

The performance of the self-consistent integrator SC--Th warrants 
further attention. We have found that the SC--Th provides full 
temperature conservation for model B. Furthermore, its results for 
$g(r)$ and $\widetilde{\kappa}_T$ are equally good to those given 
by the other integrators, and, finally, even the tracer diffusion 
results by the SC--Th are in agreement with results of other 
integration schemes. Thus, for time steps that are not too large 
(say $\Delta t \leq 0.01$), the self-consistent integrator SC--Th 
seems to provide a promising approach for studies of DPD model 
simulations. These findings contrast with those presented in 
Sect.~\ref{sec:res_modela} for the ideal gas. In the present case, the 
differentiating factor is the presence of conservative interactions. 
In model B the role of conservative forces is comparable to the random 
and dissipative contributions, suggesting that the problems in model B 
due to velocity-dependent dissipative forces are less pronounced 
than in model A. Our results support this idea. 
As demonstrated in the inset of Fig.~\ref{fig6}(b), 
$\langle \gamma(t) \rangle$ deviates only slightly 
(less than 1\%) from the desired value $\sigma^2 / 2 k_B T^{\ast}$ 
at time steps $\Delta t \leq 0.01$. For larger time increments, 
the deviations increase but remain rather small and are about 2.5\% 
around $\Delta t = 0.05$. In summary, these results demonstrate 
that, for systems where the role of random and dissipative terms 
is not dominant, the auxiliary thermostat not only minimizes
temperature deviations, but also provides a reasonable
approach for calculating dynamical quantities.

\subsection{Results for model C} 

In the preceding models, we have dealt with systems with 
soft interactions. This approach is very suitable for 
processes where the microscopic degrees of freedom do not 
matter, and where one is interested in phenomena at the 
mesoscopic level. However, there are many systems where both 
microscopic and mesoscopic properties are of interest. 
For example, the dynamics of a single polymer chain may
be studied using a model in which the polymer chain is described
in terms of Lennard-Jones interactions, while the
solvent is treated on a mesoscopic level. In this case, the role of
DPD would be to act as a thermostat and to mediate hydrodynamic
interactions, while the actual interatomic interactions within 
a polymer would be described by hard potentials. 
This approach has already proven successful in simulations 
of a system of small amphiphilic molecules, modeled by 
Lennard-Jones type interactions in conjunction with the 
DPD thermostat, \cite{soddemann_reference}  although 
no comparison of the performance of integration schemes 
was reported in that study.

To clarify the role of integrators in such cases, we examine 
this problem using model C. As described in Sect.~\ref{sec:modelc},
this model uses identical spherical particles, whose pairwise 
conservative interactions are described by a hard repulsive 
Lennard-Jones potential, while the random and dissipative 
interactions are soft. Despite its apparent simplicity, this 
approach incorporates the essential aspects required to shed 
light on this issue.

We focus on two integrators. The MD--VV integrator is chosen to 
represent an approach commonly used in molecular dynamics simulations. 
The performance of MD--VV is then compared to that of DPD--VV, 
which serves as an example of integrators designed particularly 
for DPD.

\begin{figure}[htb]
\mbox{\hspace{1.4cm}}
\psfig{file=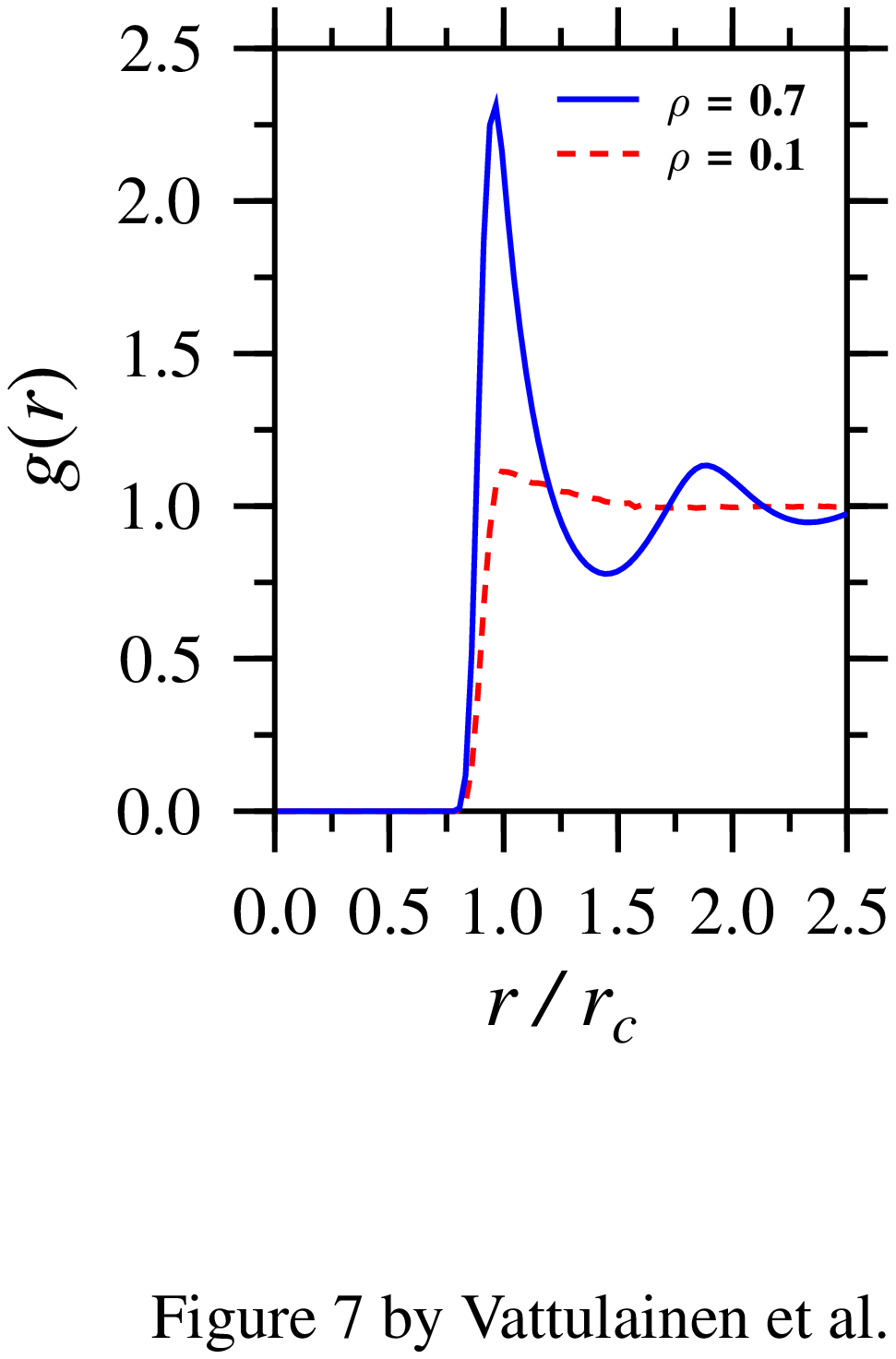,width=5.0cm,bbllx=70pt,bblly=120pt,bburx=400pt,bbury=500pt,clip=}
\caption{\label{fig7}  
Results for $g(r)$ in model C with $\sigma = 1$ using the 
integrators MD--VV and DPD--VV. Results are shown for the 
density $\rho = 0.1$ with $\Delta t = 0.01$, and for the 
density $\rho = 0.7$ with $\Delta t = 0.001$. The results 
of DPD--VV and MD--VV are essentially identical. 
}
\end{figure}

\begin{figure}[htb]
\mbox{\hspace{0.0cm}}
\psfig{file=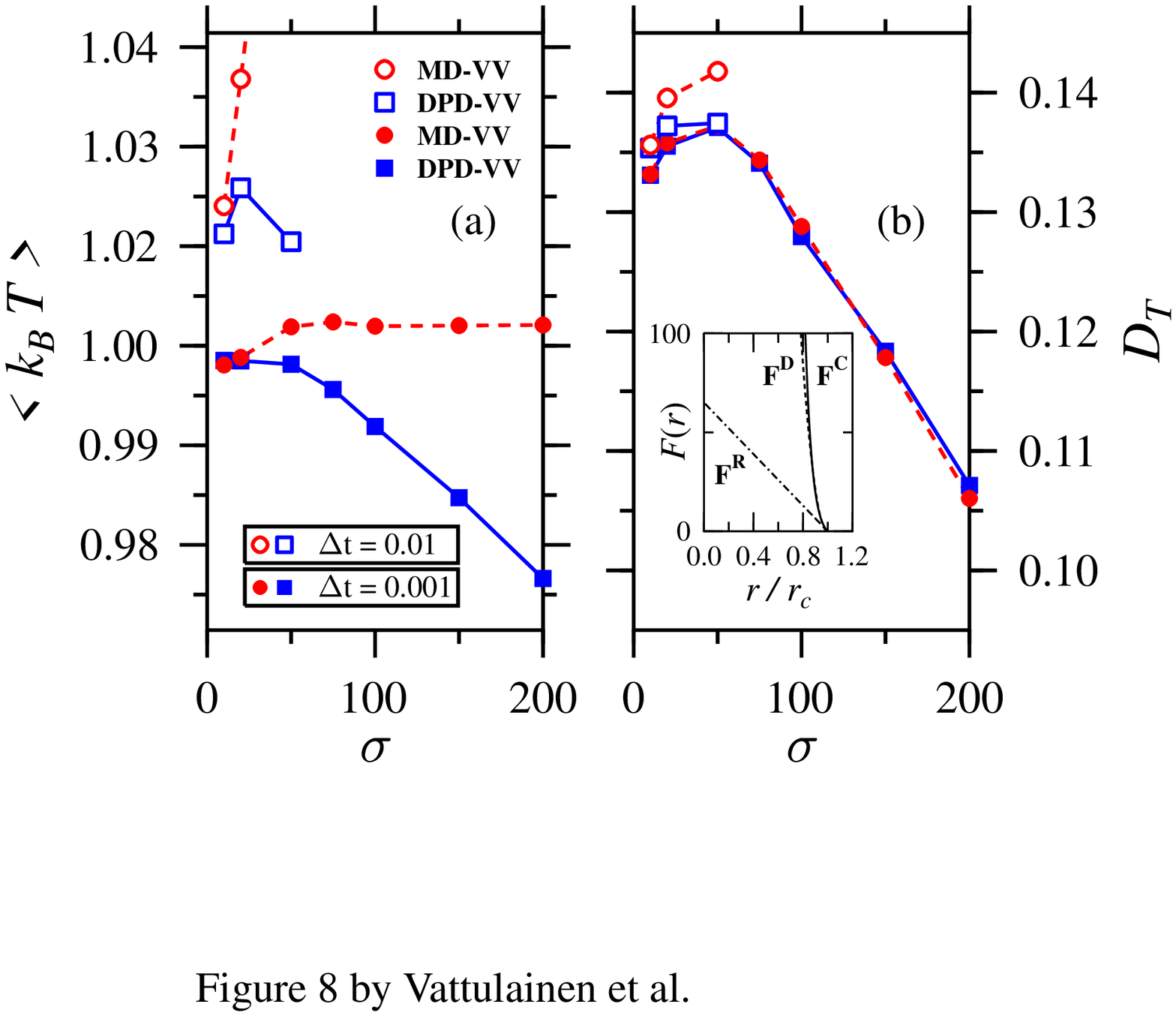,width=9.0cm,bbllx=20pt,bblly=120pt,bburx=600pt,bbury=550pt,clip=}
\caption{\label{fig8}  
Results for 
(a) the temperature $\langle k_B T \rangle$ and 
(b) the tracer diffusion coefficient $D_T$ vs. the strength 
    of the random force $\sigma$ in model C with $\rho = 0.7$. 
Results are shown for the integration schemes MD--VV and DPD--VV 
with two different time steps. For $\Delta t = 0.01$ with MD--VV, 
the system no longer remained stable beyond $\sigma = 50$. To
clarify the crossover from the regime dominated by conservative 
interactions to the regime dominated by random and dissipative 
forces, we have shown in the inset of (b) the interactions 
$F^R$ (dot-dashed), $F^D$ (dashed), and $F^C$ (full line) for 
$\sigma = 60$. For $\sigma > 60$, the dissipative force is 
steeper than the conservative one. 
}
\end{figure}

We first consider the regime predominated by conservative 
interactions. This is the case for the limit of small $\sigma$, 
where the role of random and dissipative forces is weak compared 
to that of conservative interactions. The results shown in 
Fig.~\ref{fig7} for the radial distribution function $g(r)$ 
with $\sigma = 1$ demonstrate that the system is indeed a fluid, 
and behaves in the expected manner. The radial distribution 
functions for MD--VV and DPD--VV are practically 
indistuingishable, and the same holds for the 
compressibilities extracted from the $g(r)$ data. 
Further studies in this regime revealed that the two integration 
schemes yielded rather similar results for both $\langle k_B T \rangle$ 
and $D_T$ (see results in Fig.~\ref{fig8}) as well. Differences between 
MD--VV and DPD--VV are minor at small time steps, but become 
more pronounced as $\Delta t$ is increased; around 
$\Delta t \approx 0.01$ the deviations are already significant. 
The temperature conservation shows that the artifacts 
due to MD--VV are stronger than those due to DPD--VV.
We conclude that in this regime DPD--VV performs slightly 
better than MD--VV.

The situation becomes more interesting when the random 
and dissipative forces begin to compete with conservative 
interactions. The crossover from the regime dominated by 
conservative Lennard-Jones interactions to the regime 
dominated by dissipative forces takes 
place around $\sigma \approx 60$, as is illustrated in the inset 
of Fig.~\ref{fig8}(b). Right above this threshold, it is  
evident from Fig.~\ref{fig8}(a) that $\langle k_B T \rangle$ 
starts to deviate from the desired value. In addition, as 
Fig.~\ref{fig8}(b) reveals, the tracer diffusion coefficient 
begins to decrease as soon as $\sigma$ exceeds 60. The decrease 
of $D_T$ simply reflects the fact that the dynamics are now
governed by random and dissipative forces rather than 
conservative interactions, and thus dissipation slows down 
the motion of DPD particles. Evidently there are similarities 
with Brownian motion, in which $D_T \sim 1 / \gamma$.

The fact that the dynamics in the large--$\sigma$ regime is 
controlled by random and dissipative forces leads us to expect significant
quantitative differences between MD--VV and DPD--VV, as indeed is observed.
First, in this regime the MD--VV scheme
is less stable than the DPD--VV. Second, temperature deviations 
in the case of DPD--VV are in general less pronounced as compared 
to MD--VV, and the results for the tracer diffusion behavior 
lead to similar conclusions. Thus, we conclude that, although the 
differences between MD--VV and DPD--VV are rather small, the 
DPD--VV method is more reliable for simulations in the large-$\sigma$ 
regime.

\subsection{Computational efficiency} 

In practice, the choice of an integrator is always a compromise 
between accuracy and efficiency, which in turn are related. Here, 
we briefly discuss how their mutual outcome can be optimized.

Based on Tables~\ref{table1} and \ref{table2}, it is clear that 
the efficiency of MD--VV, GW, GCC, and DPD--VV is very similar. 
The schemes GCC and DPD--VV require an additional update of 
dissipative forces, but the time it takes is negligible compared 
to the time that is required to update neighbor tables. 
Therefore, these integration schemes are approximately equally 
efficient. The self-consistent approaches, on the other hand, are 
more computer intensive. They are based on an iterative process 
to find a convergence for dissipative forces and particle 
velocities, an effort which depends on the size of the time
step. Thus, we focus on a comparison of the efficiency of the
self-consistent integration schemes to that of DPD--VV.

Using model A as a test case, we first consider SC--VV. 
As shown in Fig.~\ref{fig9}, we find that the SC--VV method 
requires three iterations per integration step to obtain 
$\langle k_B T \rangle$ with a relative accuracy of $10^{-6}$ 
at $\Delta t = 0.01$, while 20 iterations were necessary at 
$\Delta t = 0.10$ for the same accuracy. Compared to DPD--VV, 
the CPU time per integration step was increased by a factor of 
1.5 for $\Delta t = 0.10$, while it was only negligibly higher 
for the three iterations at $\Delta t = 0.01$. (The DPD--VV 
scheme corresponds to the SC--VV with ``zero iterations''.) 
Fig.~\ref{fig9} 
also shows that the number of iterations needed to obtain 
$\langle k_B T \rangle$ with a fixed accuracy increases with 
$\Delta t$, and diverges at $\Delta t \approx 0.13$ where 
the algorithm becomes unstable for the density used here.

In the case of SC--Th, the CPU time increases by a factor 
of 3 to 5 due to the extended simulation times needed to
obtain the necessary accuracy of $\langle k_B T \rangle$  
with a fluctuating $\gamma$.

These results serve to estimate the computational efficiency of 
the self-consistent integrators in systems where the conservative 
force component is very weak. In cases where the role of the 
conservative forces is more pronounced, we expect the computational 
efficiency of the self-consistent schemes to improve. This is 
due to the finding that, in models B and C, we have seen how 
temperature deviations in interacting systems are smaller than 
in the ideal gas, and thus a smaller number of iteration steps 
is expected.

We conclude that the DPD--VV is almost as fast as the MD--VV 
scheme, and the SC--VV scheme is almost as efficient as these 
simple integrators. The SC--Th scheme that includes an auxiliary 
thermostat requires considerably more time. Then it is a matter 
of taste whether the gain in temperature control is sufficient 
to justify the excess in computational effort.

\section{How tracer diffusion relates to the strength of 
         the dissipative force} 
\label{sec:diffusion}

The tracer diffusion of DPD particles has been the subject 
of various analytical studies. \cite{Gro97,Mar97,Masters99,Esp99} 
Since this topic is in part related to the present work 
(see Sect.~\ref{sec:res_modela}), we wish to discuss briefly 
the relevance of usual approximations made in describing 
the diffusion of DPD particles.

The descriptions for tracer diffusion of DPD particles 
are usually based on a few reasonable approximations. Most 
importantly, the conservative interactions are typically 
ignored and the dynamical correlations between particle 
displacements are neglected. Under these circumstances, the 
system is described by the Langevin equation (within the 
Markovian approximation), which yields the tracer diffusion 
coefficient 
\begin{equation} 
\label{eq:dtlangevin}
D_T \sim \frac{k_B T^{\ast}}{M \gamma} \, . 
\end{equation}
In practice, this expression describes the diffusion of 
a Brownian particle suspended in liquid. In this context, 
the absence of conservative interactions is justified 
since Brownian motion is driven by random forces due to 
collisions of the Brownian particle with the surrounding 
fluid particles. Neglecting dynamical correlations is also 
justified, since Brownian motion is characterized by a random 
walk in which case the velocity correlation function $\phi(t)$ 
decays exponentially in time, reflecting the lack of memory effects.

In models often studied by DPD, the case is rather different, 
however. First, DPD particles move in the presence of similar 
particles, and thus consecutive displacements of the tagged 
particle are likely to be correlated. Second, the conservative 
interactions are not irrelevant. To clarify this issue, 
i.e.,  how well Eq.~(\ref{eq:dtlangevin}) describes 
the tracer diffusion of DPD particles, we studied 
the dependence of $D_T$ on the strength of the dissipative 
force $\gamma$. To this end, we investigated models A and B 
using DPD--VV with a small $\Delta t$ (values ranging 
between $1\times 10^{-3} - 5\times 10^{-3}$).

\begin{figure}[htb]
\mbox{\hspace{0.0cm}}
\psfig{file=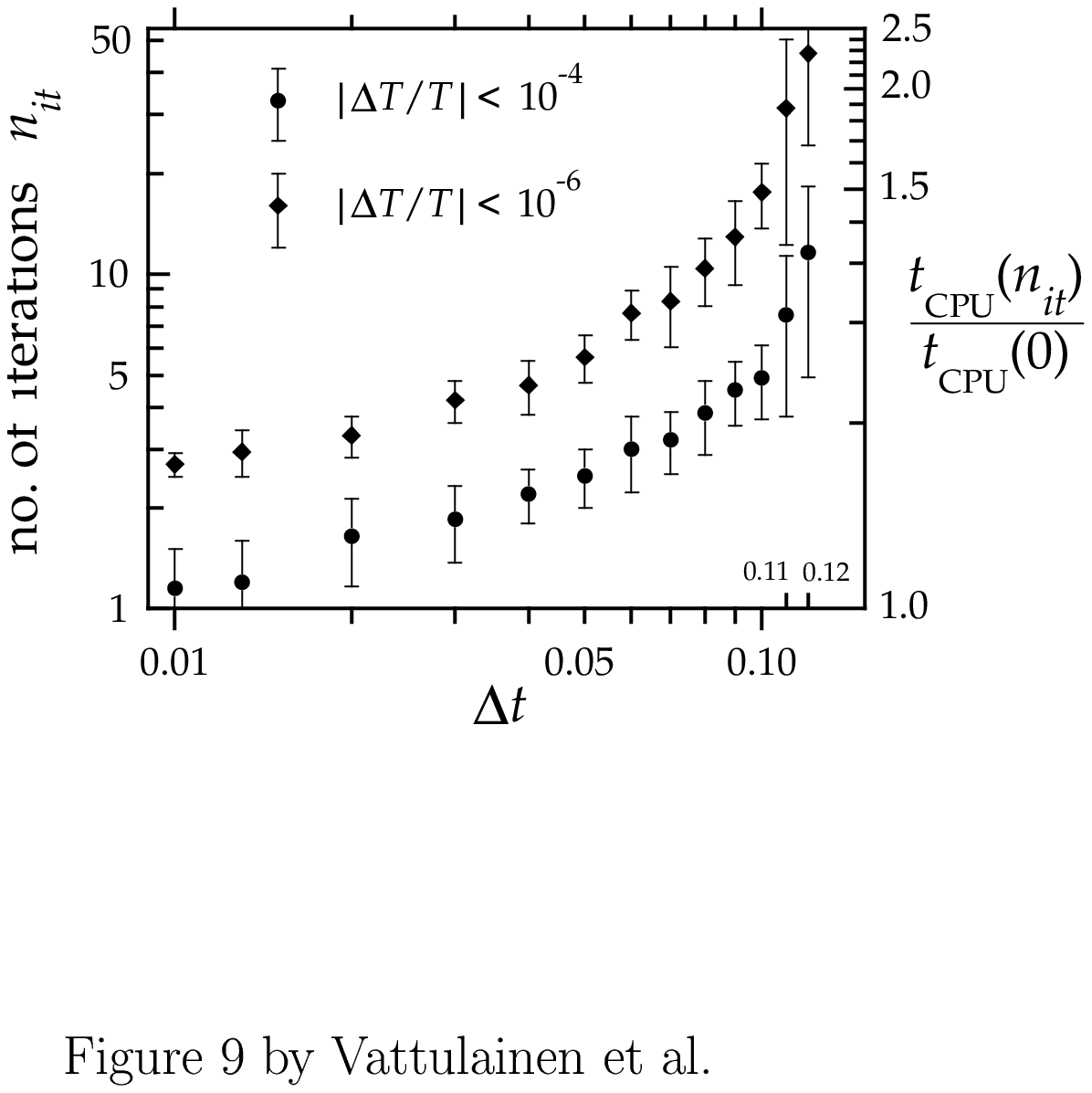,width=9.0cm,bbllx=100pt,bblly=450pt,bburx=470pt,bbury=720pt,clip=}
\caption{\label{fig9}  
Computational efficiency of the self-consistent integration 
scheme SC--VV via functional iteration. Shown here is the 
average number of iterations $n_{it}$ as a function of the 
employed time step $\Delta t$ to obtain the desired accuracy. 
The accuracy is described by the modulus of $ \Delta T / T $ 
of the instantaneous temperature, and results are shown for 
$\Delta T / T < 10^{-6}$ (solid diamonds) and 
$\Delta T / T < 10^{-4}$ (solid circles). 
The corresponding CPU time relative to the CPU time for plain 
DPD--VV (``0 iterations'') is also shown. 
}
\end{figure}

The results are presented in Fig.~\ref{fig9}. We find that 
the behavior of $D_T$ in the two models is very different 
[Fig.~\ref{fig9}(a)]. In both cases the power-law dependence 
$D_T / \langle k_B T \rangle \sim (1 / \gamma)^x$ is locally 
valid, but the exponent $x$ strongly depends on $\gamma$ 
and the strength of the conservative force $\alpha$ 
[see Fig.~\ref{fig9}(b)].

In the ideal gas ($\alpha = 0$) the motion of the DPD particles 
is fully governed by the random and dissipative forces, and so 
the exponent $x$ is approximately one at small $\gamma$. 
This behavior is expected, since then the dynamical correlations 
are very weak, as is confirmed by the exponential decay of 
$\phi(t)$ in this regime (data not shown). This is in 
agreement with recent results \cite{Masters99,Esp99} 
where $\phi(t)$ was found to decay exponentially for small 
friction. At intermediate values of $\gamma$, the power-law form 
of $D_T$ is less clear. The exponent $x$ has a minimum around 
$\gamma = 10$, and the velocity correlation function $\phi(t)$ 
decays algebraically rather than exponentially. Finally in the 
limit of large $\gamma$, $x$ tends towards one, which can be 
understood in terms of a large friction force proportional to 
$M\gamma$, and therefore this regime mimics the diffusion of DPD 
particles with a large mass. In any case, the decay of $\phi(t)$ 
is not exponential, in agreement with analytical predictions 
by Espa{\~{n}}ol and Serrano. \cite{Esp99}

In model B with finite conservative interactions, the diffusion 
at small $\gamma$ is governed by conservative interactions. This 
is best demonstrated in Fig.~\ref{fig9}(a), where $D_T$ is only 
weakly dependent on $\gamma$ in the limit of small friction. 
At intermediate values of $\gamma$, there is a crossover regime 
in which conservative and random forces compete, while at large 
$\gamma$ the diffusion behavior becomes dominated by random and 
dissipative forces. The exponent $x$ varies accordingly, reaching 
unity only in the limit of large $\gamma$.

Our aim in this work is not to focus on the diffusion properties 
of DPD model systems in detail and, thus, we do not consider this 
issue further. Nevertheless, we hope that the present results 
serve to demonstrate that the dynamics in DPD model systems 
are not similar to Brownian motion, and this dissimilarity is 
further enhanced by conservative interactions whose role can be 
significant. As regards future studies of transport properties 
of DPD fluid particles, various assumptions made in deriving 
analytical theories should therefore not be taken for granted.

\section{Summary and discussion}
\label{sec:summary}

Dissipative particle dynamics (DPD) is a very promising tool 
for future large-scale simulations of soft systems. Thus far, 
it has been applied with success to a variety of different 
problems, including studies of pressure profiles inside 
lipid bilayers, \cite{Ven99} phase behavior in surfactant 
solutions, \cite{She00} and dynamics of polymer 
chains. \cite{Spe00}

Despite its promising nature, DPD has certain practical problems 
that have to be accounted for before extensive use in future 
applications. Many of them are related to the coarse-grained 
nature of the systems studied. As described by Espa{\~{n}}ol 
and Warren, \cite{Esp95} the theoretical framework used in DPD 
leads to interparticle interactions that include a dissipative 
term, which depends on the pairwise velocities of DPD particles. 
This implies that for a proper description of the system in time, 
the dissipative forces and the particle velocities should be 
determined hand in hand in a truly self-consistent fashion. As 
shown in the present work, this issue contains various subtle 
details, but the key point is that the integration schemes often 
used in molecular dynamics simulations cannot be used in DPD 
simulations as such.

In this work, we have considered this problem through studies 
of three different model systems for a number of integration 
schemes based on the traditional velocity-Verlet approach. 
We have shown that the traditional velocity-Verlet scheme gives 
rise to pronounced artifacts in actual physical quantities such 
as the compressibility and the tracer diffusion coefficient. 
Further studies presented in this work revealed that the scale 
of these artifacts can be greatly reduced by accounting for the 
velocity dependence of dissipative forces. The simplest approach 
in this regard is to calculate the dissipative forces twice during 
a single time step, and further improvements can be obtained if 
the dissipative forces and particle velocities are determined 
together in a self-consistent fashion through a functional 
iteration process.

For cases where the remaining temperature deviations need to 
be corrected, we have proposed a self-consistent integrator
that is coupled to an auxiliary thermostat. We have discussed 
its properties through a detailed analysis in two model systems. 
We have found that this approach works well in the case of 
equilibrium quantities, whose behavior does not depend on the 
details of the dynamics. For studies of dynamical quantities 
such as diffusion, however, care must be taken to avoid misleading 
interpretations of the data. Problems may appear if 
$\langle \gamma(t) \rangle$, extracted from simulations with 
the auxiliary thermostat, deviates significantly from the dissipation 
strength $\gamma$ determined by the fluctuation-dissipation theorem. 
In practice, this situation is realized if the time step $\Delta t$ 
is relatively large and the role of conservative interactions 
is weak as compared to random and dissipative forces. However, 
if care is taken and the auxiliary thermostat is used within 
proper limits, our results show that it provides an accurate 
method to study DPD models within the NVT ensemble.

The self-consistent integrator with an auxiliary thermostat is an 
example of a scheme in which the coefficient of the dissipative 
force is not constant but fluctuates in time. Consequently, 
the average dissipative force strength $\langle \gamma(t) \rangle $
depends on the time increment $\Delta t$. Very recently, 
den Otter and Clarke suggested another approach, \cite{Ott01,Ott00} 
in which the coefficients of the random and dissipative forces 
depend on the size of the time increment $\Delta t$. 
The results presented in Ref.~\onlinecite{Ott01} 
indicate that this approach leads to good temperature control 
\cite{denotter_temp} 
compared to the GW scheme, for example, but it remains to be shown 
through thorough tests if this approach is indeed more successful 
in minimizing integrator-induced artifacts than the many other 
schemes suggested previously.

As noted in the Introduction, DPD can be thought of as Brownian 
dynamics with momentum conservation. Both methods are based on 
coarse graining the underlying microscopic systems, and in both 
cases the coarse-grained variables are replaced by random noise 
which is coupled to a dissipative friction term. Consequently,
one may question whether similar integrator-induced problems
could be faced in Brownian dynamics simulations. While we lack 
direct evidence, we feel that the problems in Brownian dynamics 
(if any) are likely less prominent compared to those in DPD. 
This idea is justified by the fact that in Brownian (Langevin) 
dynamics, the velocities of tagged particles are coupled to the 
dissipative force ($d \vec{v}_i \, / \, dt \propto -\eta \, \vec{v}_i$) 
individually for every particle. This situation is much easier 
to deal with compared to that for DPD simulations, where the 
dissipative term includes contributions from all pairs of 
particles. A thorough study of this topic would be useful.

In the present work we have examined the performance of integration 
schemes in the well-established description of dissipative particle 
dynamics, first suggested by Hoogerbrugge and Koelman \cite{Hoo92} 
and later refined by Espa{\~{n}}ol and Warren. \cite{Esp95} 
More recently, a number of related schemes have been suggested to 
shed more light on the underlying structure of DPD, \cite{Fle99} 
as well as to generalize the framework of DPD for a number of 
other hydrodynamic cases. \cite{Esp99prl} These approaches 
are numerically more complicated than the DPD considered in this 
work. Studies of the related practical issues would be very 
interesting, although they are beyond the scope of the present study.

We close this work with a brief discussion of the situations,
in which DPD-specific artifacts due to integration schemes 
are expected. To this end, we first summarize our main findings. 
We have noticed in all three model systems that various 
integrators lead to pronounced artifacts in DPD model systems, 
if random and dissipative interactions are strong compared to 
conservative interactions. On the other hand, if the system is 
governed by conservative interactions, then the artifacts have 
been found to be weaker. This suggests that one should use 
conservative interactions that are sufficiently strong to 
dominate the behavior of the model system, and let random 
and dissipative forces act only as a thermostat. 
Although this arrangement with dominating conservative forces 
is feasible in a number of cases, \cite{Dzw00a,Dzw00b} there are 
also many systems studied by DPD where random and dissipative forces 
are rather weak but {\it comparable} to conservative interactions. 

Furthermore there are processes governed by collective effects 
at large particle densities in the high friction limit, which 
based on our work can lead to integrator-induced artifacts. 
Moreover, future aims to examine soft systems of biological 
molecules in an explicit solvent lead naturally to studies of 
hybrid models, where a microscopic description for biomolecules 
is combined with a coarse-grained description for the solvent. 
In these cases there are both strong conservative interactions 
and relatively weak soft interactions, where soft interactions 
for the solvent are still subject to integrator-induced artifacts, 
and may affect the behavior of the system as whole. The overall 
picture of the role of integration schemes in specific model 
systems is therefore still incomplete, and more work is required 
to resolve these issues. \cite{Nik01}  Meanwhile, we emphasize 
that the artifacts are related to the dissipative force term, 
and therefore care should be taken in all cases where this term 
plays an important role.

\bigskip

\noindent

{\em Acknowledgements} --- 
This work has, in part, been supported by the Academy 
of Finland through its Centre of Excellence Program 
(I.V. and M.K.), and by a grant from the European Union (I.V.). 
Further support from the Alfred Kordelin Foundation and the 
Jenny and Antti Wihuri Foundation is greatly acknowledged (I.V.).

\newpage


%
%
\begin{table}[htb]
\caption{
Update scheme for a single integration step (time increment 
$\Delta t$) for various integration schemes in DPD (for acronyms 
see text). For positions and velocities at time $t$, the updated 
positions and velocities at time $t+\Delta t$ are given by the 
corresponding variables on the right-hand side of 
steps (2) and (4) below, respectively. \\
\hspace*{2em}GW($\lambda$)\,:\hfill steps (0)--(4),\,(s)\hspace*{2.8em}\\
\hspace*{2em}MD--VV $\equiv$ GW($\lambda\!=\! 1/2 $)\,:\hfill steps
(1)--(4),\,(s)\
   ${}^{\rm a}$\hspace*{2em}\\
\hspace*{2em}GCC($\lambda$)\,:\hfill steps (0)--(5),\,(s)\hspace*{2.8em}\\
\hspace*{2em}DPD--VV $\equiv$ GCC($\lambda\!=\! 1/2 $)\,:\hfill steps
(1)--(5),\,(s)\
   ${}^{\rm a}$\hspace*{2em}
}
\label{table1}
\begin{tabular}{ll}\\[-2mm]
(0) & ${\vec{v}}_i^{\,\circ}\ \longleftarrow\ \vec{v}_i\ +\
{\displaystyle \lambda\,\frac{1}{m}} \left( \vec{F}_i^C\,\Delta t +
  \vec{F}_i^D\,\Delta t +
\vec{F}_i^R\,\sqrt{\Delta t}\,\right)$\\[3mm]
(1) & $\vec{v}_i\ \longleftarrow\ \vec{v}_i\ +\
{\displaystyle \frac{1}{2}\,\frac{1}{m}} \left( \vec{F}_i^C\,\Delta t +
  \vec{F}_i^D\,\Delta t +
\vec{F}_i^R\,\sqrt{\Delta t}\,\right)$\\[4mm]
(2) & $\vec{r}_i\,\ \longleftarrow\ \vec{r}_i\hspace*{0.5mm}\ +\
\vec{v}_i\,\Delta t$\\[4mm]
(3) & Calculate\quad $\vec{F}_i^C\{\vec{r}_j\},
\quad \vec{F}_i^D\{\vec{r}_j,\,{\vec{v}}_j^{\,\circ}\},\quad
  \vec{F}_i^R\{\vec{r}_j\}$\\[3mm]
(4) & $\vec{v}_i\ \longleftarrow\ \vec{v}_i\ +\
{\displaystyle \frac{1}{2}\,\frac{1}{m}} \left( \vec{F}_i^C\,\Delta t +
  \vec{F}_i^D\,\Delta t +
\vec{F}_i^R\,\sqrt{\Delta t}\,\right)$\\[4mm]
(5) & Calculate\quad ${\vec{F}_i^D}\{\vec{r}_j,\,
{\vec{v}_j}\}$\\[1mm]
(s)\tablenotemark[2] & Calculate\quad ${k_B T} \!= {\displaystyle
  \frac{m}{3N\!-\!3}}\,
\displaystyle{\sum\limits_{i=1}^N}
\ \vec{v}_i^{\,2}$\,,\quad\ldots\\[4mm]
\end{tabular}
\tablenotetext[1]{
with substitution of \,$\vec{v}_j$\, for \,${\vec{v}}_j^{\,\circ}$\,
in step (3).
}
\tablenotetext[2]{
Sampling step [calculation of temperature $k_B T$\,, $g(r)$, \ldots]
}
\end{table}

%
%
\begin{table}[htb]
\caption{
Update scheme for the two self-consistent integrators without
(SC--VV) and with (SC--Th) the auxiliary thermostat [steps (i)--(iii)].
The self-consistency loop is over steps (4b) and (5) as indicated.
For positions and velocities at time $t$, the updated positions and
velocities at time $t+\Delta t$ are given by the corresponding variables 
on the right-hand side of steps (2) and (4b) below (after the last 
iteration of the self-consistency loop), respectively. 
The desired temperature is $k_B T^{\ast}$. Initialization: 
$\eta = 0, \gamma = {\sigma}^2 / (2k_B T^{\ast})$, and $k_B T$
is calculated from the initial velocity distribution.
}
\label{table2}
\begin{tabular}{lll}\\[-2mm]
{} & (i) & ${\bf \dot{\eta}}\ \hspace*{0.47em}\longleftarrow\ C\,(k_B T -
   k_B T^{\displaystyle \ast})$\\[3mm]
{} & (ii) & ${\eta}\ \hspace*{0.47em}\longleftarrow\ \eta\ +\ {\bf \dot{\eta}}\,
   \Delta t$\\[2mm]
{} & (iii) & $\gamma\ \hspace*{0.37em}\longleftarrow\ {\displaystyle
   \frac{\sigma^2}{2\,k_B T^{\displaystyle \ast}}}\,(1 +
   \eta\,\Delta t)$\\[3mm]
{} & (1) & $\vec{v}_i\ \longleftarrow\ \vec{v}_i\ +\
   {\displaystyle \frac{1}{2}\,\frac{1}{m}} \left( \vec{F}_i^C\,\Delta t +
   \vec{F}_i^D\,\Delta t + \vec{F}_i^R\,\sqrt{\Delta t}\,\right)$\\[4mm]
{} & (2) & $\vec{r}_i\,\ \longleftarrow\ \vec{r}_i\hspace*{0.5mm}\ +\
   \vec{v}_i\,\Delta t$\\[4mm]
{} & (3) & Calculate\quad $\vec{F}_i^C\{\vec{r}_j\}, \quad \vec{F}_i^D
   \{\vec{r}_j,\,\vec{v}_j\},\quad \vec{F}_i^R\{\vec{r}_j\}$\\[3mm]
{} & (4a) & $\widehat{\vec{v}}_i\ \longleftarrow\ \vec{v}_i\ +\
   {\displaystyle \frac{1}{2}\,\frac{1}{m}} \left\{ \vec{F}_i^C\,\Delta t +
   \vec{F}_i^R\,\sqrt{\Delta t}\,\right\}$\\[3mm]
   {} & (4b) & $\vec{v}_i\ \longleftarrow\ \widehat{\vec{v}}_i\ +\
   {\displaystyle \frac{1}{2}\,\,\frac{1}{m}}\,
   \vec{F}_i^D\,\Delta t$\\[-4mm]
\setlength{\unitlength}{0.1em}
\begin{picture}(6,30)
 \put(0,3){\line(1,0){6}}
 \put(0,3){\line(0,1){23}}
 \put(0,26){\vector(1,0){6}}
\end{picture}&
(5) & Calculate\quad ${\vec{F}_i^D}\{\vec{r}_j,\,
{\vec{v}_j}\}$\\[1mm]
{} & (s) & Calculate\quad ${k_B T} \!= {\displaystyle \frac{m}{3N\!-\!3}}\,
\displaystyle{\sum\limits_{i=1}^N}
\ \vec{v}_i^{\,2}$\,,\quad\ldots\\[4mm]
\end{tabular}
\end{table}
%



\begin{figure}[htb]
\mbox{\hspace{0.0cm}}
\psfig{file=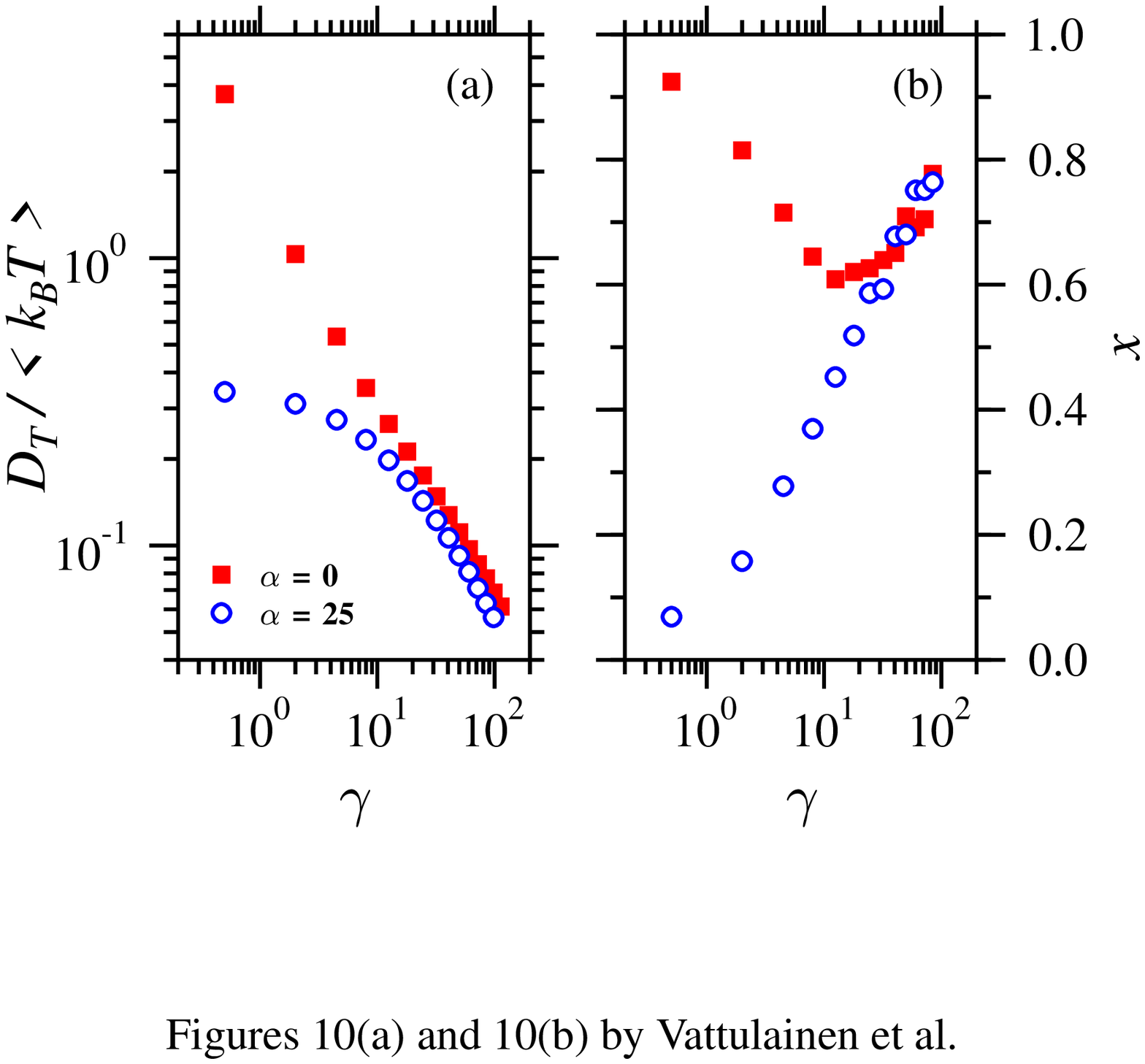,width=9.0cm,bbllx=20pt,bblly=120pt,bburx=600pt,bbury=550pt,clip=}
\caption{\label{fig10}  
(a) Diffusion results for $D_T / \langle k_B T \rangle$ vs. the 
strength of the dissipative force $\gamma$ in DPD simulations 
for models A ($\alpha = 0$) and B ($\alpha = 25$). 
(b) Based on the results in (a) and using the ansatz 
$D_T / \langle k_B T \rangle \sim \gamma^{-x}$, here is 
shown the resulting exponent $x$ as a function of $\gamma$. 
The results shown here have been calculated by DPD--VV 
with small $\Delta t$ (ranging from $1\times 10^{-3}$ to 
$5\times 10^{-3}$) such that temperature deviations 
in all cases are very minor. 
}
\end{figure}


\begin{thebibliography}{250}

\bibitem{Fre96}
D. Frenkel and B. Smit,   
{\it Understanding Molecular Simulation: From Algorithms to Applications} 
(Academic Press, San Diego, 1996). 

\bibitem{Lad93}
A. J. C. Ladd, 
Phys. Rev. Lett. {\bf 70}, 1339 (1993); 
J. Fluid Mech. {\bf 271}, 285 (1994); 
J. Fluid Mech. {\bf 271}, 311 (1994). 

\bibitem{Mur98} 
M. Murat and K. Kremer, 
J. Chem. Phys. {\bf 108}, 4340 (1998).

\bibitem{Mal99}
A. Malevanets and R. Kapral, 
J. Chem. Phys. {\bf 110}, 8605 (1999); 
J. Chem. Phys. {\bf 112}, 7260 (2000). 

\bibitem{Hoo92} 
P. J. Hoogerbrugge and J. M. V. A. Koelman, 
Europhys. Lett. {\bf 19}, 155 (1992). 

\bibitem{Esp95} 
P. Espa{\~{n}}ol and P. Warren, 
Europhys. Lett. {\bf 30}, 191 (1995). 

\bibitem{Gro97} 
R. D. Groot and P. B. Warren, 
J. Chem. Phys. {\bf 107}, 4423 (1997). 

\bibitem{War98} 
P. B. Warren, 
Curr. Opin. Colloid. Interf. Sci. {\bf 3}, 620 (1998). 

\bibitem{Gro99} 
R. D. Groot, T. J. Madden, and D. J. Tildesley, 
J. Chem. Phys. {\bf 110}, 9739 (1999). 

\bibitem{Ven99} 
M. Venturoli and B. Smit, 
PhysChemComm {\bf 10} (article 10) (1999). 

\bibitem{Nov98} 
K. E. Novik and P. V. Coveney, 
J. Chem. Phys. {\bf 109}, 7667 (1998). 

\bibitem{Pag98} 
I. Pagonabarraga, M. H. J. Hagen, and D. Frenkel, 
Europhys. Lett. {\bf 42}, 377 (1998). 

\bibitem{Gib99} 
J. B. Gibson, K. Chen, and S. Chynoweth, 
Int. J. Mod. Phys. C {\bf 10}, 241 (1999). 

\bibitem{Ott01} 
W. K. den Otter and J. H. R. Clarke, 
Europhys. Lett. {\bf 53}, 426 (2001). 

\bibitem{Ahl99}
P. Ahlrichs and B. D\"unweg, 
J. Chem. Phys. {\bf 111}, 8225 (1999). 

\bibitem{Mal00}
A. Malevanets and J. Yeomans,  
Europhys. Lett. {\bf 52}, 231 (2000). 

\bibitem{Bes00} 
G. Besold, I. Vattulainen, M. Karttunen, and J. M. Polson, 
Phys. Rev. E {\bf 62}, R7611 (2000).

\bibitem{For95} 
B. M. Forrest and U. W. Suter, 
J. Chem. Phys. {\bf 102}, 7256 (1995). 

\bibitem{Fle99} 
E. G. Flekk{\o}y and P. V. Coveney, 
Phys. Rev. Lett. {\bf 83}, 1775 (1999); 
E. G. Flekk{\o}y, P. V. Coveney, and G. De Fabritiis, 
Phys. Rev. E {\bf 62}, 2140 (2000).  

\bibitem{Gar83} 
C. W. Gardiner, 
{\it Handbook of Stochastic Methods} 
(Springer-Verlag, Berlin, 1983). 

\bibitem{dunweg91a} 
B. D{\"u}nweg and W. Paul,
Int. J. Mod. Phys. C, {\bf 2}, 817 (1991). 

\bibitem{Lec88} 
P. L'Ecuyer, 
Commun. ACM {\bf 31}, 742 (1988). 

\bibitem{Pre92} 
W. H. Press, S. A. Teukolsky, W. T. Vetterling, B. P. Flannery, 
{\it Numerical Recipes in Fortran, The Art of Scientific 
Computing}, 2nd Edition 
(Cambridge University Press, Cambridge, 1992) pp. 271--273. 

\bibitem{Vat01} 
I. Vattulainen, 
submitted (2001). 

\bibitem{Tuc00}
M. E. Tuckerman and G. J. Martyna, 
J. Phys. Chem. B {\bf 104}, 159 (2000). 

\bibitem{Ver67} 
L. Verlet, 
Phys. Rev. {\bf 159}, 98 (1967). 

\bibitem{All93} 
M. P. Allen and D. J. Tildesley, 
{\it Computer Simulation of Liquids} 
(Oxford University Press, Oxford, 1993). 

\bibitem{Mar95} 
G. J. Martyna and M. E. Tuckerman, 
J. Chem. Phys. {\bf 102}, 8071 (1995). 

\bibitem{Thi99} 
J. M. Thijssen, 
{\it Computational Physics} 
(Cambridge University Press, Cambridge, 1999). 

\bibitem{Boo80} 
J. P. Boon and S. Yip, 
{\it Molecular Hydrodynamics} 
(Dover, New York, 1980). 

\bibitem{compr} 
For $\protect\Delta t \protect\gtrsim 0.1$, the positive 
and negative deviations from $g(r) = 1$ {\it accidentally} 
almost cancel each other. 

\bibitem{Han86} 
J.-P. Hansen and I. R. McDonald, 
{\it Theory of Simple Liquids}, 2nd edition 
(Academic Press, San Diego, 1986). 

\bibitem{soddemann_reference}
Th. Soddemann, B. D\"unweg, and K. Kremer 
(unpublished). 


\bibitem{Mar97} 
C. A. Marsh, G. Backx, and M. H. Ernst, 
Europhys. Lett. {\bf 38}, 411 (1997); 
Phys. Rev. E {\bf 56}, 1676 (1997). 

\bibitem{Masters99} 
A. J. Masters and P. B. Warren, 
Europhys. Lett. {\bf 48}, 1 (1999). 

\bibitem{Esp99} 
P. Espa{\~{n}}ol and M. Serrano, 
Phys. Rev. E {\bf 59}, 6340 (1999). 

\bibitem{She00} 
J. C. Shelley and M. Y. Shelley, 
Curr. Opin. Colloid Interface Sci. {\bf 5}, 101 (2000). 

\bibitem{Spe00} 
N. A. Spenley, 
Europhys. Lett. {\bf 49}, 534 (2000). 

\bibitem{Ott00} 
W. K. den Otter and J. H. R. Clarke, 
Int. J. Mod. Phys. C {\bf 11}, 1179 (2000). 

\bibitem{denotter_temp} 
Den Otter and Clarke have pointed out that there is no 
unique way to define the temperature of the system, 
\protect\cite{Ott01,Ott00}  and that the different definitions 
(in terms of the kinetic energy or the configurational 
temperature, for example) may not be completely consistent 
in the context of DPD. 

\bibitem{Esp99prl}
P. Espa{\~{n}}ol, M. Serrano, and H. C. \"Ottinger, 
Phys. Rev. Lett. {\bf 83}, 4542 (1999); 
M. Serrano and P. Espa{\~{n}}ol, 
Phys. Rev. E {\bf 64}, 046115 (2001). 

\bibitem{Dzw00a} 
W. Dzwinel and D. A. Yuen, 
J. Colloid Interface Sci. {\bf 225}, 179 (2000). 

\bibitem{Dzw00b} 
W. Dzwinel and D. A. Yuen, 
Int. J. Mod. Phys. C {\bf 11}, 1 (2000). 

\bibitem{Nik01} 
P. Nikunen, M. Karttunen, and I. Vattulainen 
(unpublished). 


\end{thebibliography}
\end{document}